\documentclass[fleqn,usenatbib]{mnras}

\usepackage{newtxtext,newtxmath}
\usepackage[T1]{fontenc}

\DeclareRobustCommand{\VAN}[3]{#2}
\let\VANthebibliography\thebibliography
\def\thebibliography{\DeclareRobustCommand{\VAN}[3]{##3}\VANthebibliography}

\usepackage{graphicx}	% Including figure files
\usepackage{amsmath}	% Advanced maths commands

\usepackage{amssymb}	% Extra maths symbols
\usepackage{cleveref}   % cross-cref
\usepackage{ulem}

\usepackage{multirow}
\usepackage{pdflscape}

\title[Earth Microlensing Zone]{Earth through the looking glass: how frequently are we detected by other civilisations through photometric microlensing?}

\author[S. Suphapolthaworn et al.]{S. Suphapolthaworn,$^{1}$
S. Awiphan,$^{2}$\thanks{E-mail: supachai@narit.or.th}
T. Chatchadanoraset,$^{3}$
E. Kerins,$^{4}$
D. Specht,$^{4}$
N. Nakharutai,$^{5}$ \newauthor
S. Komonjinda,$^{6}$
and A.C. Robin$^{7}$
\\
$^{1}$ Department of Cosmosciences, Graduate School of Science, Hokkaido University, Sapporo, Hokkaido 060-0810, Japan\\
$^{2}$ National Astronomical Research Institute of Thailand (Public Organization), 260 Moo 4, Donkaew, Mae Rim, Chiang Mai 50180, Thailand\\
$^{3}$ Chiang Mai University Demonstration School, Chiang Mai 50200, Thailand\\
$^{4}$ Jodrell Bank Centre for Astrophysics, University of Manchester, Oxford Road, Manchester M13 9PL, UK\\
$^{5}$ Data Science Research Center, Department of Statistics, Faculty of Science, Chiang Mai University, Chiang Mai 50200, Thailand\\
$^{6}$ Department of Physics and Materials Science, Faculty of Science, Chiang Mai University, Chiang Mai 50200, Thailand\\
$^{7}$ Institut Utinam, CNRS UMR 6213, Univ. Bourgogne Franche-Comté, OSU THETA, Observatoire de Besançon, BP 1615 25010 Besançon Cedex, France
}

\date{Accepted XXX. Received YYY; in original form ZZZ}

\pubyear{2022}

\begin{document}
\label{firstpage}
\pagerange{\pageref{firstpage}--\pageref{lastpage}}
\maketitle

% Abstract of the paper
\begin{abstract}
Microlensing is proving to be one of the best techniques to detect distant, low-mass planets around the most common stars in the Galaxy. In principle, Earth's microlensing signal could offer the chance for other technological civilisations to find the Earth across Galactic distances. We consider the photometric microlensing signal of Earth to other potential technological civilisations and dub the regions of our Galaxy from which Earth's photometric microlensing signal is most readily observable as the ``Earth Microlensing Zone'' (EMZ). The EMZ can be thought of as the microlensing analogue of the Earth Transit Zone (ETZ) from where observers see Earth transit the Sun. Just as for the ETZ, the EMZ could represent a game-theoretic Schelling point for targeted searches for extra-terrestrial intelligence (SETI). To compute the EMZ, we use the \textit{Gaia} DR2 catalogue with magnitude $G<20$ to generate Earth microlensing probability and detection rate maps to other observers. Whilst our Solar system is a multi-planet system, we show that Earth's photometric microlensing signature is almost always well approximated by a binary lens assumption. We then show that the Earth is in fact well-hidden to observers with technology comparable to our own. Specifically, even if observers are located around every \textit{Gaia} DR2 star with $G<20$, we expect photometric microlensing signatures from the Earth to be observable on average only tens per year by any of them. In addition, the EMZs overlap with the ETZ near the Galactic centre which could be the main areas for future SETI searches.

\end{abstract}

\begin{keywords} gravitational lensing: micro -- astrobiology -- planets and satellites: general -- stars: statistics
\end{keywords}

%%%%%%%%%%%%%%%%%%%%%%%%%%%%%%%%%%%%%%%%%%%%%%%%%%

%%%%%%%%%%%%%%%%% BODY OF PAPER %%%%%%%%%%%%%%%%%%

\section{Introduction}

To date, almost 5000 exoplanets\footnote{From: \texttt{https://exoplanetarchive.ipac.caltech.edu/} (NASA exoplanet archive).} have been discovered by a number of surveys (e.g. \textit{Kepler} \citep{bor2005}, TESS \citep{ric2014}, WASP \citep{pol2006,smi2014}, HATNet \citep{bak2004,bak2009}, KELT \citep{pep2007}, NGTS \citep{whe2018}, MASCARA \citep{sne2012}, MOA \citep{bon2001,sum2003}, OGLE \citep{uda1993} and KMTNet \citep{lee2015}) and thousands will be found with on-going and up-coming programs (e.g. PLATO \citep{rau2014}, NASA \textit{Nancy Grace Roman Space Telescope} (hereafter \textit{Roman}, \citet{spe2013,spe2015}) and \textit{Euclid} \citep{pen2013,mcd2014}). Although thousands of exoplanets have been found, most of them are located in the \textit{Kepler} field that covers only 0.25 percent of the sky. To provide better exoplanetary statistics, transit surveys are being planned or are underway that cover larger sky areas, including all-sky surveys. Additionally, astrometric and microlensing exoplanet samples are expected to be grown rapidly from space-based surveys.

As we learn more about exoplanets, more scientists are starting to consider more carefully the search for extra-terrestrial intelligence (SETI). Many survey strategies have been conducted by SETI projects, including targeting specific sky areas or systems \citep{tur2003a,tur2003b,sie2013,enr2017,mar2018,pin2019,pri2020,kal2021}. Sun-like stars with planets in their habitable zones are one of the main targets for SETI searches \citep{tur2003a,mar2018}. The mere existence of the Earth, a rocky planet within the habitable zone of the Sun, could provide motivation for extra-terrestrial observers to send signals to or to listen for signals from the Earth. At present, the majority of confirmed exoplanets have been detected by the transit technique$^1$ that can  be used to detect planetary atmospheres, and it can be potentially applied to detect even biosignatures or industrial pollution \citep{lin2014}. However, the transit technique is limited by the orbital inclination of the planets. As a result, there is only a narrow strip of sky centred on the Ecliptic plane where extraterrestrial observers could see the Earth as a transiting exoplanet; the ``Earth Transit Zone'' (ETZ). The ETZ is therefore a promising region for SETI technosignature searches \citep{fil1988,hel2016,wel2018,ker2021,kal2020,she2020,kal2021}.

The microlensing technique is based on the gravitational lens effect induced when a foreground (lensing) star or planet passes in front of a background (source) star. The source light can become multiply imaged and distorted, resulting in an overall magnification of the source. If the lensing star has a planet orbiting around it, the planet can perturb the light, leading to additional spikes in the light curve \citep{mao1991}. At present, over 130 microlensing exoplanets have been discovered by three main microlensing programs: MOA \citep{bon2001,sum2003}, OGLE \citep{uda1993} and KMTNet \citep{lee2015}, and potentially thousands of microlensing exoplanets could be discovered with the NASA \textit{Roman} telescope \citep{pen2019}) and by the ESA \textit{Euclid} mission \citep{pen2013}. Microlensing events have also been proposed for targeted SETI searches \citep{rah2016}. 

Since we have only recently gained the capability of detecting extra-solar planets, it is natural to consider how these techniques could be used by other civilisations to detect us. Recently \cite{ker2021} has advocated the strategy of Mutual Detectability as a game-theory based approach to targeted SETI. Targeting SETI searches towards systems that may have a good view of the Earth is an example of a ``Schelling Point'' strategy, an optimal strategy within cooperative games involving non-communicating participants \citep{sch1960,wri2017}.  In the context of SETI, one major advantage of microlensing over other known exoplanet detection methods is that it is long range. Over 130 confirmed exoplanets have been detected using microlensing, typically involving planetary systems located between 3-7~kpc from us in the direction of the Galactic bulge. By contrast, other known detection methods are typically confined to around 1~kpc from the Sun. If technological civilisations are rare in our Galaxy then the best chance of the Earth being detected may come from a method capable of long range (Galactic scale) utility. By considering how detectable the Earth may be to other civilizations via microlensing, we can gauge the extent to which distant observers should be considered within a Mutual Detectability strategy.

In this paper we focus specifically on photometric microlensing since it is the simplest manifestation of microlensing. Nevertheless, we note that, in principle, astrometric microlensing offers potentially a larger detection cross section. There are good reasons to believe that some parts of our Galaxy may be more conducive to life than others. However, for the purposes of our calculations, we will take an agnostic view and assume that technological civilisations have equal chance to be located around any star anywhere in the Galaxy. With this assumption we can define the ``Earth microlensing zone'' (EMZ) as the regions of the sky from which observers may most likely see Earth's microlensing signal. This likelihood may be higher either because there are many stars in that region (hence more sites for observers to exist) or because observers in the regions have access to many background sources against which Earth's signal may be seen. The EMZ are then, in some sense, the microlensing analogue of the ETZ. Note that the EMZ is a region with a higher statistical probability of containing an observer that detects the Earth through microlensing, all other factors being equal. Unlike the ETZ, it is not a region within which any observer is guaranteed to see a signal from the Earth. Furthermore, whilst transit observations have the advantage of periodicity (and therefore the possibility of repeated follow-up observations) microlensing events are transient for any given observer.

The extent and location of the EMZ can be evaluated from the sky density of potential observers weighted by microlensing probability or detection rate. The \textit{Gaia} mission Data Release 2 (\textit{Gaia} DR2) \citep{Gaia2018} contains the positions and optical brightness of $\sim$1.7 billion stars. Around 1.3 billion stars in the catalogue have precise measurements of distance and proper motion calculated by \cite{bai2018}. Therefore, in this work, the \textit{Gaia} DR2 data are used to construct the EMZ for stars with \textit{Gaia} magnitude $G < 20$. This magnitude limit is decided from the completeness of \textit{Gaia} DR2 data.

Our work is organised as follows. In \Cref{sec:micro_properties}, our calculation and assumptions are described. In order to ensure that the Sun-Earth binary lens caustic approximation can be used to calculate the EMZ, the lensing effects of major bodies in the Solar system are considered in \Cref{sec:solarsystem}. The observer/source selection criteria of the \textit{Gaia} DR2 catalogue and the map resolution are provided in \Cref{sec:simulation}. In \Cref{sec:result}, the Earth microlensing zone is constructed from the distributions of microlensing observers. Finally, \Cref{sec:conclusion} discusses and summarises the work.

\section{Microlensing properties}
\label{sec:micro_properties}

To construct the EMZ, we consider here the signal produced by the Earth acting as a lens that passes in front of some background stars from the point of view of some extrasolar observers. From the Earth's viewpoint, the extraterrestrial observer and the background star are located at their respective antipodes from each other. We divide the sky into small regions (pixels) and consider the Earth microlensing probability for observers within each pixel, taking into account their lines of sight as well as the magnitude and distance of stars at their antipode pixels that act as potential sources.

We consider three main microlensing properties: (i) microlensing probability, which in this case refers to the probability that any observer within a sky pixel sees a microlensing signal at any given time due to the Earth acting as the lens; (ii) the average caustic crossing time of Earth's microlensing signal averaged over all observer and source locations for a given sky pixel; and (iii) the rate at which any observer within a sky pixel may observe an Earth-induced microlensing event.

To make our calculations, it is necessary to adopt some strong assumptions. All of our calculations are normalised to an assumption of one active observer population per star. The maps can be easily rescaled to any other provided number of observer populations. We also assume that all stars have equal likelihood of hosting  observers, irrespective of stellar host types, ages or locations.

\subsection{Microlensing probabilities}
\label{sec:prob}
In order to calculate the probability of microlensing by the Earth ($P$) along a line of sight at any given time, the microlensing optical depth ($\tau$) which is the number of ongoing events at a given time is needed. For a single lens microlensing, the optical depth is a fraction of the sky covered by the Einstein radius. The Einstein radius can be written as 

\begin{equation}
\theta_E=\sqrt{\frac{4GM(D_\textup{s}-D_\textup{l})}{c^2D_\textup{s} D_\textup{l}}} \ , 
\end{equation}
where $G$ is the gravitational constant, $M$ is the lens mass, $c$ is the speed of light, $D_\textup{s}$ is the distance between an observer and a source, and $D_\textup{l}$ is the distance between an observer and a lens. The optical depth of foreground lenses within the solid angle ($\Omega$) for an observer ($\textup{o}$) can be written as:

\begin{equation}\label{eq:tau}
\tau_{s,\textup{o}}=\sum_{\textup{s}=0}^{N_\textup{s}} \frac{\pi \theta_{E,\textup{o},\textup{s}}^2}{\Omega_\textup{s}} \ ,
\end{equation}
where the subscription $\textup{s}$ indicates the source, and $N_\textup{s}$ is the number of sources.

However, as the Earth orbiting the Sun, the extrasolar observers could observe the microlensing event produced by the Earth as a multiple-lens microlensing event. We will show in \Cref{sec:solarsystem} that, for photometric microlensing, it is almost always safe to ignore the effect of the other Solar system planets when considering Earth's microlensing signature. In this case we can use a binary lens assumption and consider the binary caustic areas for the effective microlensing cross-section area instead of the single lens Einstein radius. The central caustic area in unit of $\theta_{E}$ can be approximated as,

\begin{equation}\label{eq:central}
A_{c} \simeq \frac{1}{2}\Delta\xi_{\textup{c,c}}\Delta \eta_{\textup{c,c}} \ ,
\end{equation}
where $\Delta\xi_{\textup{c,c}}$ and $\Delta \eta_{\textup{c,c}}$ are the central caustic sizes defined in Equations 11 and 12 of \citet{chu2005}, respectively. However, the total caustic area is dominated by the areas of the two planetary caustics which their total areas in unit of $\theta_{E}$ can be approximated as,

\begin{equation}\label{eq:planet}
    A_{p} \simeq \begin{cases}
               \frac{\Delta\xi_{\textup{c,p,l}}}{2}\frac{\Delta \eta_{\textup{c,p,l}}}{2}\ ,\hspace{10mm} s<1 \\
               \frac{1}{2}\Delta\xi_{\textup{c,p,g}}\Delta \eta_{\textup{c,p,g}}\ ,\hspace{5.5mm} s\geq1 \\
        \end{cases}\ ,
\end{equation}
where $\frac{\Delta\xi_{\textup{c,p,l}}}{2}$, $\frac{\Delta \eta_{\textup{c,p,l}}}{2}$, $\Delta\xi_{\textup{c,p,g}}$ and $\Delta\eta_{\textup{c,p,g}}$ are the size of planetary caustics which are defined in Equations 18, 13, 8 and 9 of \citet{han2006}, respectively.

For the Galactic microlensing, the Earth generally locates inside the Sun's Einstein radius ($s<1$, where $s$ is the Earth-Sun separation in the unit of the Sun's Einstein radius). As the Earth orbiting the Sun, the separation between the Earth and the Sun depends on the Earth location and the direction of the observers. The separation can be written as

\begin{equation}
s = \sqrt{\left (a_e \sin \phi \right )^2+\left (b_e \cos \phi \right )^2} \ ,
\end{equation}
where
\begin{equation}
a_e = \frac{1 \text{AU}}{D_l}  \ ,
\end{equation}
\begin{equation}
b_e = \frac{1 \text{AU}}{D_l}\sin(i_e) \ ,
\end{equation}
$\phi$ is the Earth's orbital phase, and $i_e$ is an ecliptic latitude. Therefore, for an observer, the time average total caustic area over a year, when the Earth completes a full orbit around the Sun of a binary lens can be written as,

\begin{equation}\label{eq:tau_b}
\tau_{b,\textup{o}}=\sum_{\textup{s}=0}^{N_\textup{s}} \frac{\int_{0}^{2\pi}\left (A_{c,\textup{o},\textup{s}}(s)+A_{p,\textup{o},\textup{s}}(s) \right )d\phi}{2\pi}\frac{\theta_{E,\textup{o},\textup{s}}^2}{\Omega_\textup{s}} \ .
\end{equation}

However, there is discrepancy between the analytic approximation and the exact caustic size near $s=1$ \citep{chu2005,han2006}. Although the cases are very rare for the Galactic microlensing events, the approximations will overapproximate the size of the caustic, which affects the microlensing probabilities calculation. In order to avoid the large optical depth values of the system near $s=1$, the separation is set to be,

\begin{equation}
    s = \begin{cases}
               0.99,\hspace{30mm} 0.99\leq s<1.00 \\
               1.01,\hspace{30mm} 1.00\leq s<1.01 \\
               s,\hspace{34.5mm} \textup{otherwise} \\
        \end{cases}\ .
\end{equation}

The obtained optical depth can be used to calculate the probability of microlensing given by,

\begin{equation}
P = 1-e^{-\tau} \ .
\end{equation}
In the case of Galactic microlensing, the optical depth is small. Therefore, the probability of microlensing can be approximated as $P\simeq\tau$. For a small area of observers, the total probability that those observers can witness microlensing events is

\begin{equation}\label{eq:prob_micro}
P = 1-\left (\prod_{\textup{o}=0}^{N_\textup{o}}(1-P_\textup{o})\right ) = 1 - \left (1-\sum_{\textup{o}=0}^{N_\textup{o}} P_\textup{o} + ... \right ) \approx \sum_{\textup{o}=0}^{N_\textup{\textup{o}}} P_\textup{o} \approx \sum_{\textup{o}=0}^{N_\textup{o}} \tau_\textup{o}\ ,
\end{equation}
where $N_\textup{o}$ is the number of observers. Throughout this work, microlensing probability is defined as $P\approx \sum_{\textup{o}=0}^{N_\textup{o}} \tau_{\textup{o}}$. 

\subsection{Average caustic crossing time}

The caustic crossing time is the time that source passes through the caustic of the lensing object,

\begin{equation}
t_E = \textup{min}\left ( \sqrt{\frac{\int_{0}^{2\pi}\left (A_{c,\textup{o},\textup{s}}(s)+A_{p,\textup{o},\textup{s}}(s) \right )d\phi}{2}}, 1 \right ) \frac{\theta_E}{\mu} \ ,
\end{equation}
where $\mu$ is the lens–source pair-wise relative proper motion. For an observer at the Earth, proper motion of the extrasolar observer, $\mu_\textup{o}$, and proper motion of source, $\mu_\textup{s}$, can be obtained. The Earth-source relative proper motion for an extrasolar observer can be written as,

\begin{equation}
\vec{\mathbf{\mu}}=\vec{\mathbf{\mu_\textup{s}}}+\vec{\mathbf{\mu_\textup{o}}} \ .
\end{equation}
The average caustic crossing time for an observer is
\begin{equation}
\left \langle t_E \right \rangle_\textup{o} = \frac{\sum_{\textup{s}=0}^{N_\textup{s}} t_{E,\textup{o},\textup{s}}}{N_\textup{s}} \ .
\end{equation}

\subsection{Microlensing discovery rates}

Assuming microlensing events occur when the lens-source angular separation comes within the caustic area of the lens-source pair. The Earth microlensing discovery rate for an observer is

\begin{equation}
\Gamma_\textup{o} = \frac{2P}{\pi \left \langle t_E \right \rangle_\textup{o}} \ .
\end{equation}
The total discovery rate for observers in a small area can be calculated as follows:

\begin{equation}
\Gamma = \sum_{\textup{o}=0}^{N_\textup{o}} \Gamma_{\textup{o}} \ .
\end{equation}

In order to calculate average caustic crossing time for all observers, the total microlensing probabilities and discovery rates in a small area are used as follows:

\begin{equation}
\left \langle t_E \right \rangle = \frac{2P}{\pi\Gamma} \ .
\end{equation}

\section{The Solar system caustic network}
\label{sec:solarsystem}

In a microlensing scenario where the Earth produces a caustic crossing event, it is important to consider the effect of other objects in the Solar system, predominantly the influence of Jupiter and Saturn and to a lesser extent, Uranus, Neptune, Mars and Venus. The addition of these extra planets to the lensing system adds the possibility of requiring more complicated lensing models than the binary lens caustic in \Cref{sec:prob}, such as a trinary or higher order lensing model. Such models are difficult to fit to photometry due to the higher dimensionality of the parameter space and the additional degeneracies introduced by the complicated, asymmetric caustic network. As such, it is important to analyse the significance of these effects and to what extent the binary model is sufficient to describe a typical Earth induced caustic crossing event. To this end, an eight-fold lensing system was simulated, including the Sun, Venus, Earth, Mars, Jupiter, Saturn, Uranus and Neptune.

A major difficulty in computing the magnification map for an eight-fold lens is overcoming the expensive operation of evaluating the valid image positions, which involves evaluating an $n^2 + 1=65^{\rm th}$ order polynomial \citep{dan2015}. As we are interested in finding every root to the lens equation, finding the eigenvalues of the corresponding companion matrix is necessary. This process itself is of order $\mathcal{O}(n^3)$, making the combined time complexity of solving the lens equation for an n$^\textup{th}$ order lens $\mathcal{O}(n^5)$. To overcome this computational intractability, a ray-casting algorithm was used instead, which has a time complexity of order $\mathcal{O}(n)$. 

\begin{figure*}
    \centering
    \includegraphics[width=\textwidth]{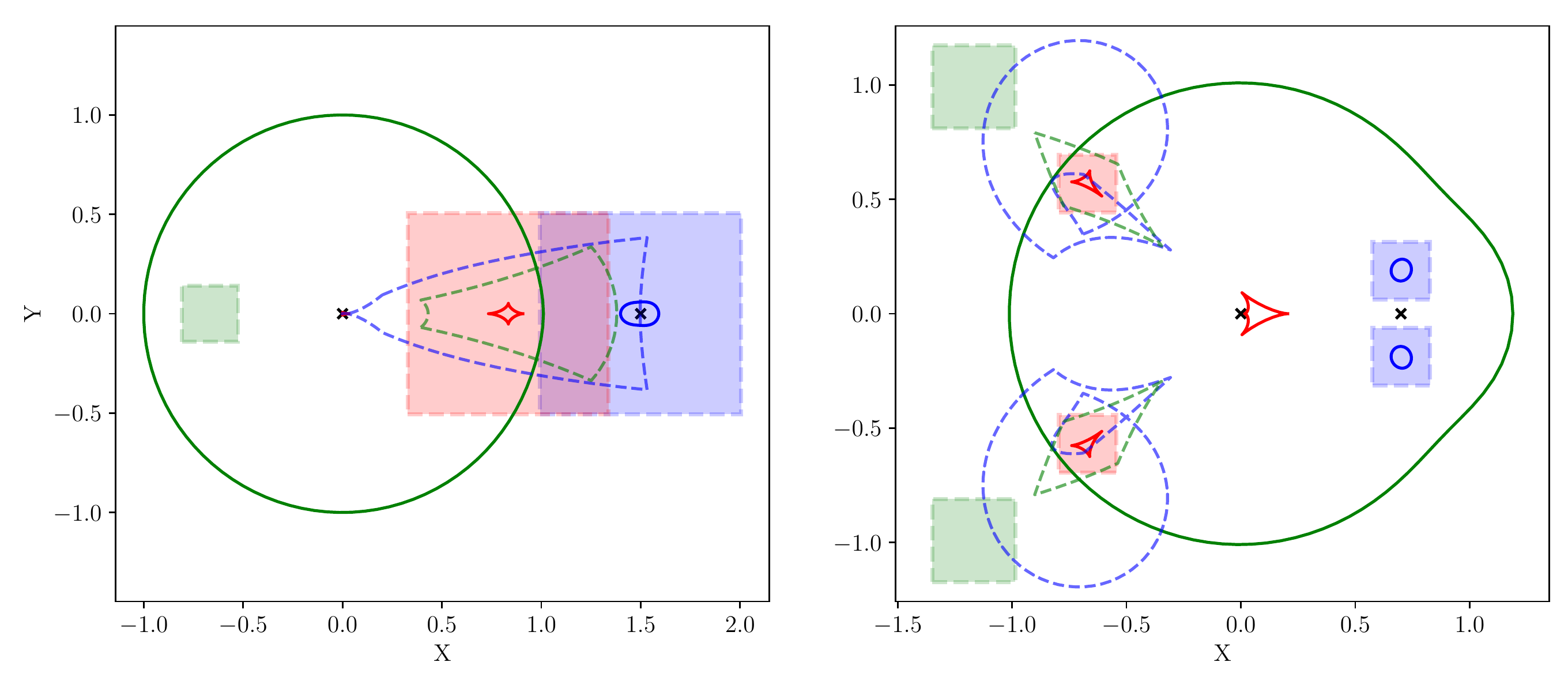}
    \caption{The ray-casting emission and receiver planes, superimposed over the microlensing critical boundaries and caustics for wide (left, $q=0.005$, $s=1.5$) and close (right, $q=0.07$, $s=0.7$) binary lensing. The solid green and blue lines represent the critical boundaries of the primary and planetary lenses, respectively. The solid red lines represent the caustics, with the planetary caustic(s) covered by the receiver plane(s), indicated with a red square. The emission planes over the planetary critical boundaries are shown as blue squares, while the emission planes used for the magnification contribution from the primary lens are shown as green squares. In the source plane, the un-lensed locations of the emission planes are indicated with dashed blue and green lines, which overlap with the receiver plane and fully cover the planetary caustic(s). The lens locations are indicated with black crosses.}
    \label{Fig:Raycasting}
\end{figure*}

This process works in two stages: calculating the appropriate locations and sizes for the ray-cast emission planes and then the same for the ray-cast receiver planes. Emission planes are square regions in the lens plane, placed over each lens' critical boundary, through which the rays are cast. An additional emission plane is placed over a location in the lens plane corresponding to the background magnification induced by the primary lens in the region around the caustic. The ray-cast receiver planes are placed over the central location of the target caustic, the location of which is refined over ten steps with a Newton-Raphson iteration, using a starting location for the caustic given by the equivalent location from binary lensing, as derived in \citet{han2006}. This initial guess allows the iteration to converge assuming the mass ratios of the higher order lenses relative to the primary lens are small, on the order of $q < 10^{-2}$. The sizes of each plane are calculated to fully cover the target caustic and are based on the caustic dimensions reported in \citet{han2006}. For the emission planes over critical boundaries and the receiver planes, the size $d_1$ is given by

\begin{equation}
    d_1 = \begin{cases}
               \frac{16 \sqrt{q}}{\sqrt{s^2 - 1}}\ ,\hspace{27.5mm} \textup{wide} \\
               8 \sqrt{q} \frac{k_0 (s^2 + 1) - s^2}{k_0 s^2} \cos(\theta / 2)\ ,\hspace{3mm}\textup{close} \\
        \end{cases}\ ,
\end{equation}
where the parameters $k_0$ and $\theta$ are given by

\begin{equation}
    k_0 = \sqrt{\frac{ \cos(\theta) - \sqrt{s^4 - \sin(\theta)^2} }{s^2 - s^{-2}}} \ ,
\end{equation}

\begin{equation}
    \theta = \pi - \arcsin \Big( s^2 \frac{\sqrt{3}}{2} \Big)\ .
\end{equation}
For the emission plane over the primary lens, a different size $d_2$ was used,

\begin{equation}
    d_2 = \begin{cases}
               4\sqrt{q}\ , \hspace{6mm} \textup{wide} \\
               4s^3\sqrt{q}\ , \hspace{3mm} \textup{close}
          \end{cases} \ .
\end{equation}
The final configuration of the emission and receiver planes is shown in \Cref{Fig:Raycasting}, which shows an example of the raycasting plane locations in the case of binary lensing; for illustration purposes, a comparatively large binary parameter $q$ is used and plane sizes are reduced.

Once the locations of each plane had been determined, rays at a density of 20 per pixel and a resolution of $500\times500$ pixels from the emission plane are cast, with their locations represented by the complex number $z = x+iy$ in the image plane transformed using the N-fold lens equation to their corresponding location $w$ in the source plane via

\begin{equation}
    w = z - \sum_{n=1}^{N}\frac{\mu_n}{\Bar{z}-\Bar{z}_n} \ ,
\end{equation}
where $\mu_j = M_n/M_{\rm tot}$ is a lens mass normalised to the total lens system mass and $z_n$ is the corresponding lens location. Once $w$ has been evaluated, the magnification contribution for each ray cast is added to the receiver plane at the pixel corresponding to that location, with a scaling factor of $(d_2/d_1)^2$ applied to rays from the primary lens emission plane, to account for the differing raycast densities.

A final clean-up stage was then applied, as only the portion of the receiver plane covered by the projections of both emission planes was usable. To that end, a rectangular mask was applied to the receiver, covering the maximal overlap area of both emission planes, with a low resolution background covering the whole lens system out to $\pm 10\theta_{\rm E}$ applied underneath.

\begin{figure*}
    \centering
    \includegraphics[width=\textwidth]{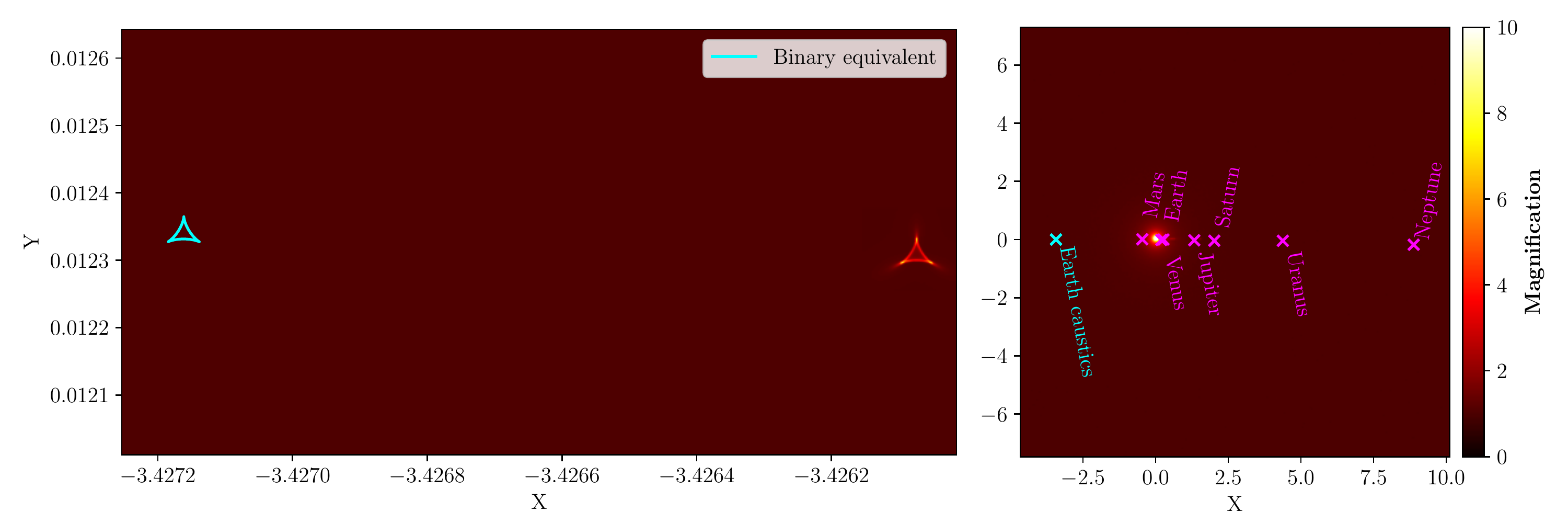}
    \caption{The magnification map for the eight-fold lens is shown, zoomed out, on the right. The locations of each of the lenses are indicated, which are based on realistic locations at the epoch MJD=59507. On the left, the magnification map centred on one of the Earth's planetary caustics (the location of which is indicated on the right with a cyan marker); two caustics are shown, with the red (right) caustic representing the magnification map from the eight-fold lens and the cyan (left) caustic representing the equivalent caustic shape and location for a binary Earth-Sun lensing configuration. The binary caustic produced by the Earth is generally shifted but not distorted when the effect of the other planets is included. The result of photometric microlensing signals due to the Earth can therefore safely assume the binary lens approximation.}
    \label{Fig:caustic_network}
\end{figure*}

Two regions were simulated; firstly, a magnification map of the whole Solar system was simulated for various planetary alignments and secondly, a zoomed in region centred on the planetary caustics produced by the Earth. Realistic binary parameter values for the normalised angular separation, $s$, and planetary to host mass ratio, $q$, were used for each planetary lens, with the Earth's separation set to $s=0.3$. \Cref{Fig:caustic_network} shows the result of the raytracing algorithm for both the full Solar system and the specific Earth planetary caustics and also shows the equivalent location and shape of the Earth-Sun binary caustic. The positions of the planetary lenses were projected along a line of sight in Ecliptic coordinates of $(l,b)=(270^\circ,0^\circ)$, which is close to the Galactic centre, a favoured mutual detectability region. While it is clear that the introduction of other planetary lenses induces an offset between the eight-fold caustic and the binary equivalent, the shapes and sizes of the caustics are almost exactly the same. Using a metric $\delta$ of the relative difference in caustic area given by

\begin{equation}
    \delta = \frac{A_{\rm binary} - A_{\rm 8-fold}}{A_{\rm 8-fold}} \ ,
\end{equation}
where $A_{\rm binary}$ is the area of Earth's planetary caustic in the binary scenario and $A_{\rm 8-fold}$ is the equivalent for the 8-fold lens scenario. For the configuration shown in \Cref{Fig:caustic_network}, this metric gives $\delta=-0.0003$, suggesting that a typical alignment of the Solar system towards the EMZ does not produce a caustic network that warrants a trinary or higher order lens model when considering an Earth caustic crossing event.

\section{Simulating Earth as microlensing planet}\label{sec:simulation}

\subsection{\textit{Gaia} DR2 catalogue}\label{sec:gaia}

In order to calculate microlensing properties, stars' positions, brightness, and proper motions are needed. Therefore, the \textit{Gaia} DR2 catalogue, which contains $\sim$1.3 billion sources with positions and proper motions \citep{Gaia2018}, is used. For sources brighter than $G=20$, the catalogue provides proper motion with uncertainty of 1.2 mas yr$^{-1}$. The source distances are obtained from the estimated distance catalogue of \cite{bai2018}, which estimates the distance in parsec from the \textit{Gaia} parallaxes.

\begin{figure}
\centering
\includegraphics[width=0.5\textwidth]{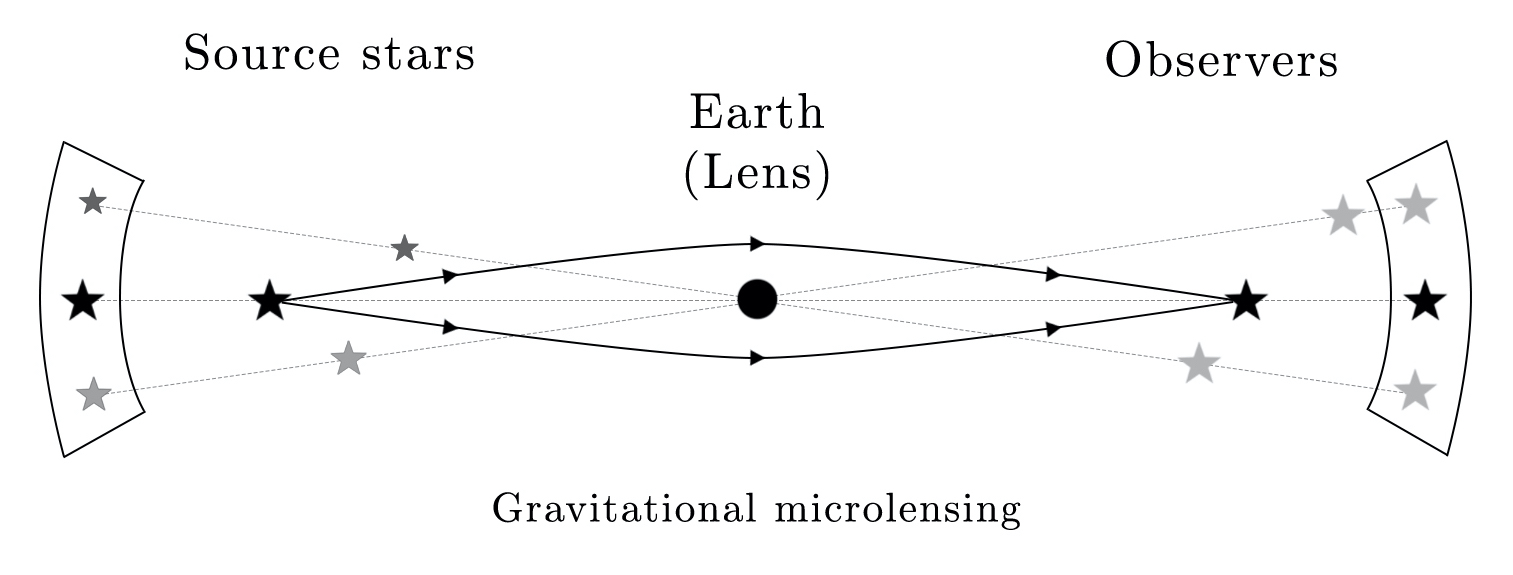}
\caption{An illustration of a gravitational microlensing event and observer-source pairings for this study. In order to calculate microlensing properties of each observer, an observer is selected from stars within one HEALPix area and paired with every star in its antipode HEALPix area.}
\label{Fig:Cartoon}
\end{figure}

The \textit{Gaia}'s \texttt{source\_id} is encoded using the nested Hierarchical Equal Area isoLatitude Pixelation (HEALPix) \citep{gor1999} scheme at level 12, which divides the sky into $\simeq$200 million pixels with resolution $\sim$0.014$^\circ$. In this work, the HEALPix scheme is used to divide the sky area into small pieces to calculate microlensing properties. The stars in divided areas are assigned to be observers and background sources for the calculation. For the observers in such an area, they could observe the Solar system with the Earth passing in front of stars in their antipode HEALPix area. Assuming that an observer can see the Earth and all background stars in the antipode HEALPix area located within the same line of sight, the observer is paired with every background star within that HEALPix area (see \Cref{Fig:Cartoon}). Note that only the Earth is assumed to be the companion in the binary lens as discussed in \Cref{sec:solarsystem}. Thus, the Earth's mass is implemented as the mass of the companion lens for all calculations.

The \textit{Gaia} $G$-band mean flux is adopted in order to define the source brightness. The observers could see the background stars fainter than an observer at the Earth, as they are located further away. To make the calculations more accurate, extinction along the path is added. The $G$-band mean magnitude that could be observed by each observer is
\begin{equation}
G_\textup{obs} = G_\textup{Earth} - 5\log_{10} \left (\frac{D_\textup{s}-D_\textup{o}}{D_\textup{s}} \right ) + A_\textup{o} \ ,
\end{equation}
where $G_\textup{Earth}$ is $G$-band mean magnitude of the object observed from the Earth. $D_\textup{s}$ and $D_\textup{o}$ are the distances from the Earth to the source and to the observer, respectively. $A_\textup{o}$ is the total extinction along the path between the observer and the Earth.

The Galactic interstellar dust model of \cite{lal2019} is used to calculate the total extinction along that path. The extinction model of \cite{lal2019} has a resolution of 25 pc. In order to prevent repeatedly counting the same extinction bin in the area located further away from the Earth, the total extinction calculation process taking place at every 50 pc interval towards the direction of each HEALPix pixel is considered in this work. In addition, if a star locates further than the limits of this model, the remaining distance would be treated as the same extinction of the last point within the limit of the model. For simplicity, if the apparent brightness of the source at the observer is fainter than $G=20$, it is not included in the calculation. Note that the blending effect is neglected in this work.

\subsection{Map resolutions}

The level of HEALPix scheme designates the resolution index of the map. The index determines the total number of pixels that the sky is divided into. The number of pixels in different levels of HEALPix scheme, $N$, can be calculated using the following equation

\begin{equation}
N = 12 \times {\left (2^\textup{level} \right )}^2 \ .
\end{equation}
For instance, HEALPix level 6 divides the sky into 49 152 pixels with a resolution of $\sim$0.916$^\circ$, while level 12 divides the sky into $\simeq$200 million pixels with a resolution of $\sim$0.014$^\circ$. Thus, a HEALPix level 6 pixel contains 4096 pixels of HEALPix level 12 within it.

The \textit{Gaia}'s \texttt{source\_id} is encoded under the HEALPix scheme at level 12. The location of a star in terms of the HEALPix pixel number, $\textup{N}$, can be obtained using $\textup{N}~=~\texttt{source\_id}\://\:\left(2^{35}\times 4^{(12-\textup{level})}\right )$, where $//$ designates a floor division. Therefore, the HEALPix pixel number of stars can be calculated based on the HEALPix scheme at any level. Throughout this work, only the HEALPix scheme at levels 6 and 12 are applied.

\begin{figure*}
\centering
\includegraphics[width=0.45\textwidth]{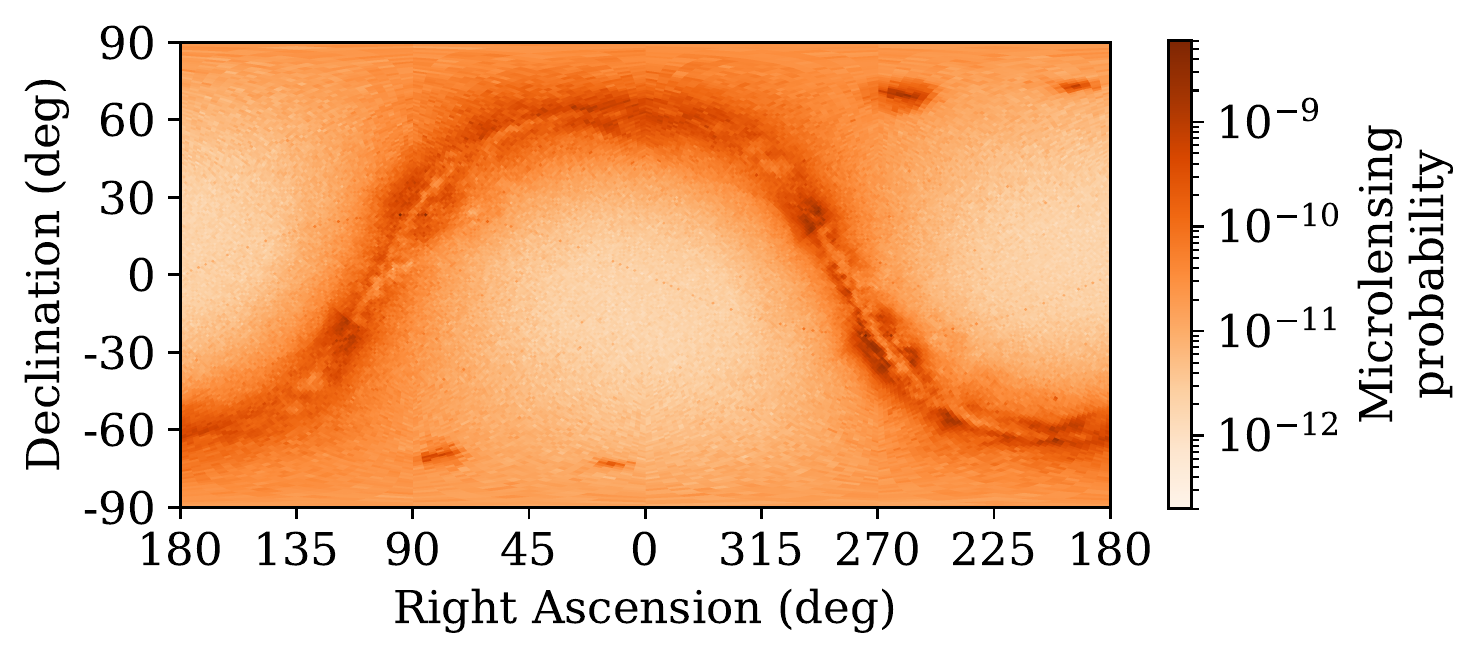} 
\includegraphics[width=0.45\textwidth]{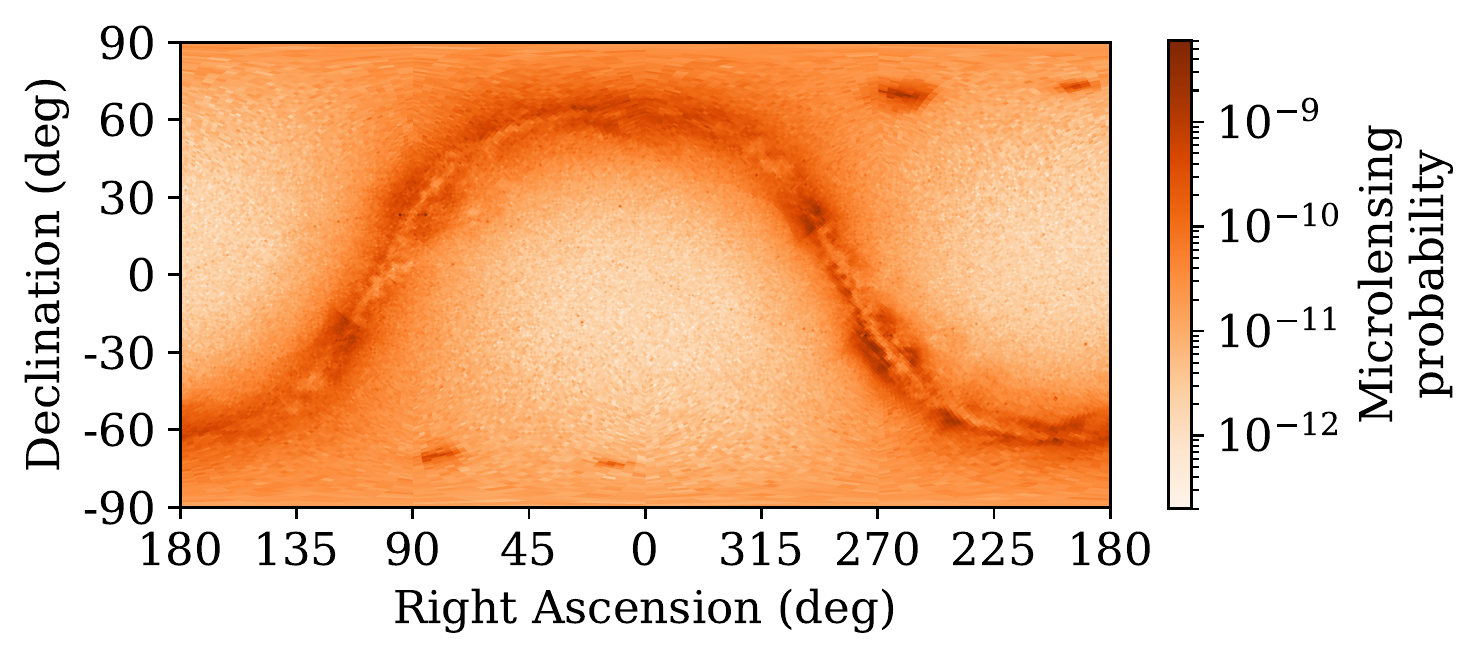} \\
\includegraphics[width=0.45\textwidth]{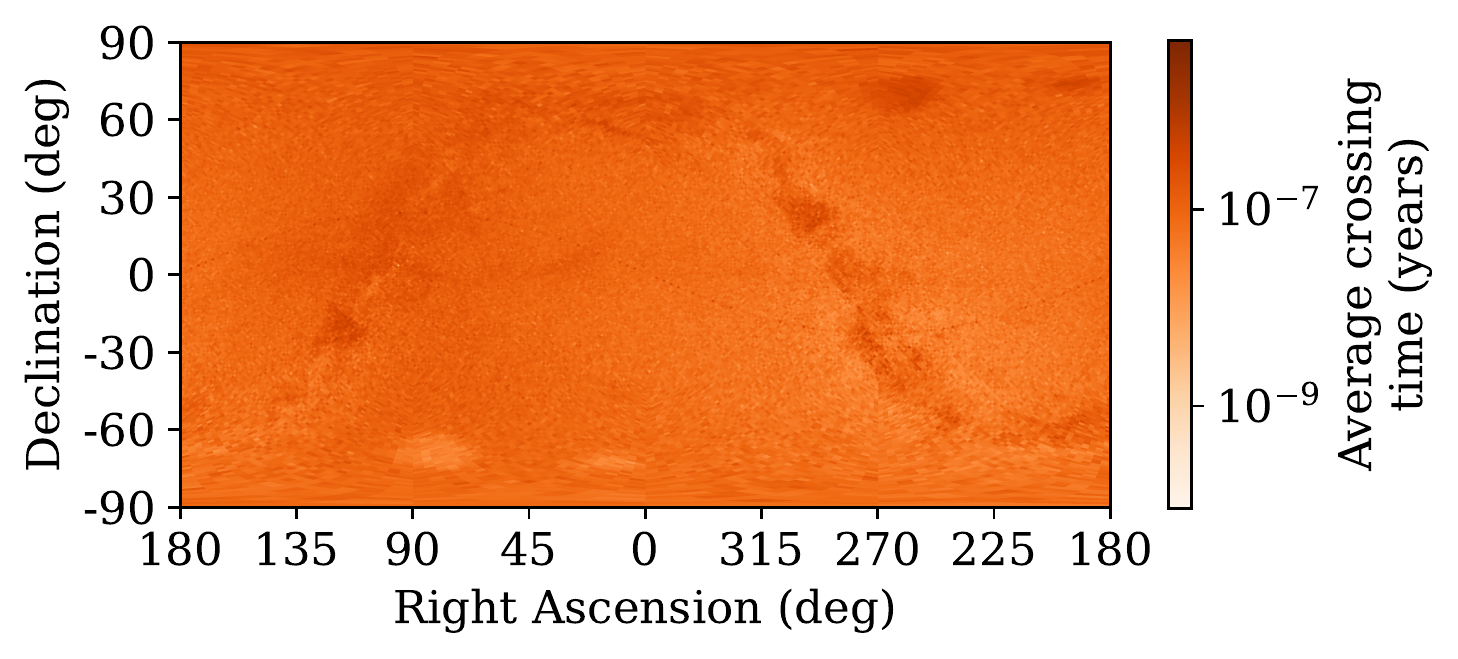}
\includegraphics[width=0.45\textwidth]{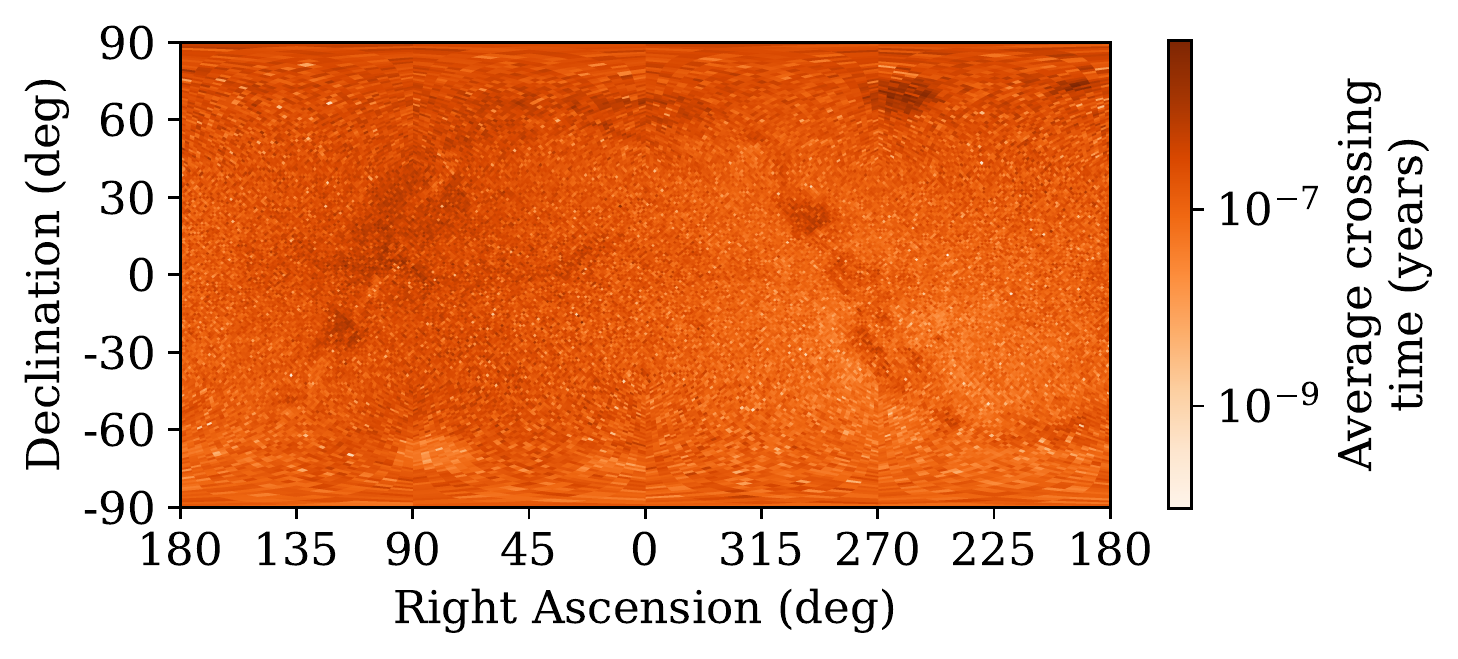}\\
\includegraphics[width=0.45\textwidth]{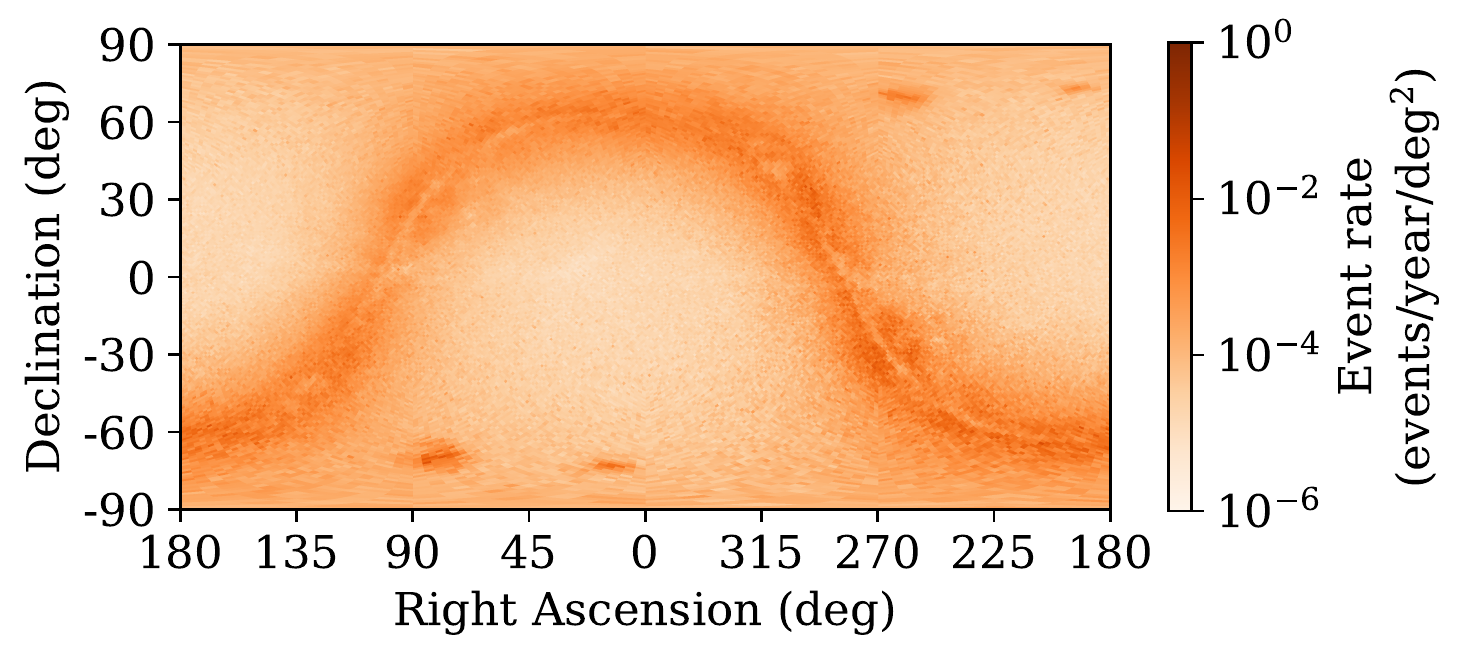}
\includegraphics[width=0.45\textwidth]{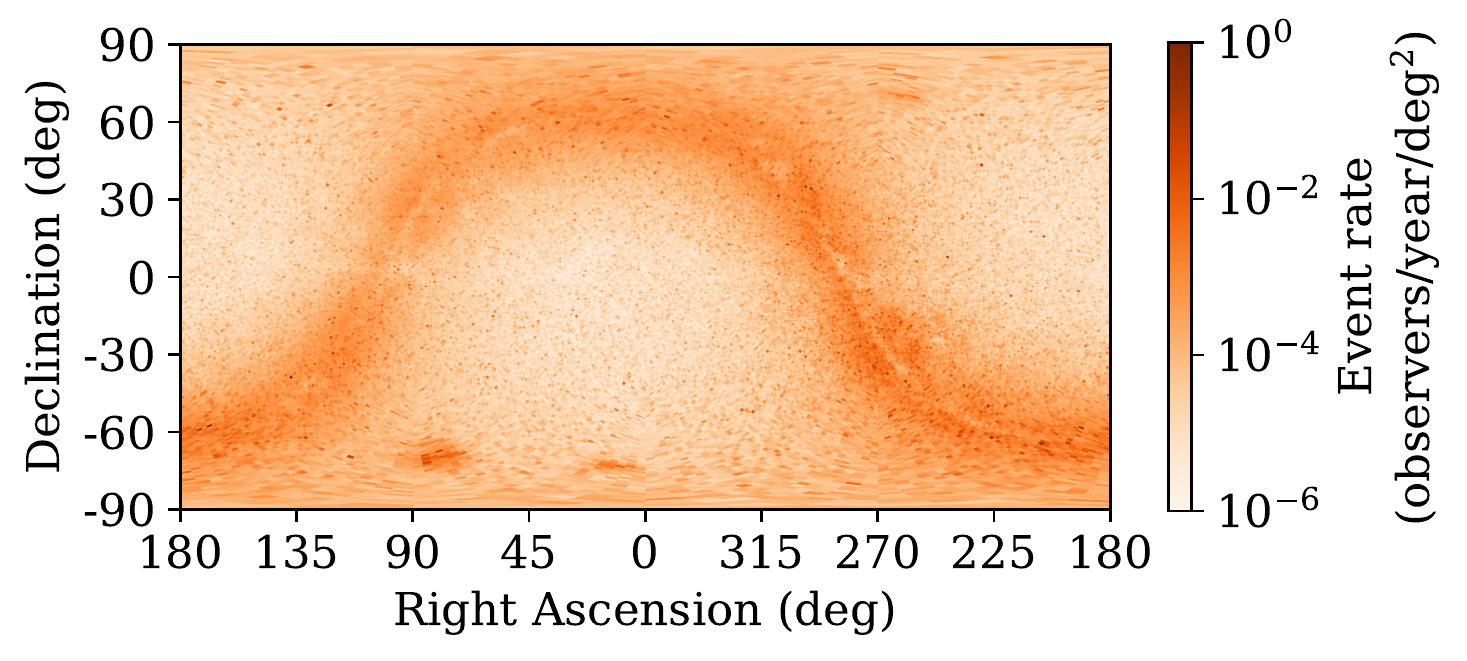} \\

\caption{Microlensing probability (Top), average caustic crossing time (Middle) and Earth discovery rate (Bottom) maps with HEALPix scheme at level 6 (Left) and level 12 (Right). The maps with HEALPix level 12 are mapped as level 6 pixel by summing the values from every level 12 HEALPix located within the level 6 HEALPix area.}
\label{Fig:MapHP6HP12}
\end{figure*}

Initially, the microlensing probability, average caustic crossing time and the Earth discovery rates for an observer at HEALPix level 12 are calculated. However, the size of HEALPix level 12 is too small to be illustrated in an all-sky map and the position of the stars heavily relies on the time when \textit{Gaia} observed the stars. Furthermore, there are a number of pixels at HEALPix level 12 that do not possess either source, observers or both in the pixels. For visualisation and to make sure that there is at least one source-observer pair in each pixel, the values at HEALPix level 12 are mapped as HEALPix level 6 pixels by summing over all level 12 values within the area of level 6 pixels. The maps of both HEALPix scheme level 6 and 12 are shown in \Cref{Fig:MapHP6HP12}. 

In general, there is no major difference between microlensing parameters calculated from HEALPix level 6 and HEALPix level 12 schemes. The microlensing probabilities from the two schemes are similar in areas between $10^\circ$ above and below the Galactic plane. Moreover, the average caustic crossing time in the HEALPix level 6 scheme in this area and the Galactic anti-centre is slightly shorter than that in HEALPix level 12. As a result, the event rate in HEALPix level 6 is higher compared to HEALPix level 12 in this area.

HEALPix scheme is defined by the area with a constant solid angle. Thus, the solid angle of HEALPix level 12 is smaller than that of HEALPix level 6. From the difference in the size of solid angle, each observer in HEALPix level 12 has fewer sources than the observers in HEALPix level 6. Although the lower number of sources can be seen in the level 12 maps, the number of sources does not affect the microlensing properties because all the properties are scaled with the solid angle of the HEALPix. However, the HEALPix scheme is defined by the observer on the Earth. For extra-solar observers, their apparent solid angle sizes depend on the distance between them and the source. Therefore, in this work, the apparent solid angle of each observer and source are scaled with the inverse square of the distance between observer and source ($\Omega_s \propto \left ( D_{\textup{Earth-source}}/D_{\textup{s}} \right )^{2}$). The observers located near the Earth have an apparent solid angle with a size similar to the apparent solid angle of the HEALPix, while the observers located further away have apparent solid angles smaller than the HEALPix solid angle. After applying the correction in the calculation, there is no major discrepancy between the map of level 6 and level 12.

As the calculations based on HEALPix at level 12 is more dependent on the time of observation than HEALPix at level 6 and there is no major difference between the HEALPix scheme at level 12 and level 6, the HEALPix scheme at level 6 is used in the rest of this study.

\section{Earth detectability}\label{sec:result}

\subsection{Microlensing probability, average caustic crossing time and Earth discovery rate distributions}\label{sec:properties_map}

The microlensing probability, average caustic crossing time and Earth discovery rate maps from the calculations done at HEALPix scheme level 6 are shown in \Cref{Fig:MapHP6HP12} and \Cref{Tab:MAPHP6HP12}. The microlensing probability has a strong relation with the numbers of observers and sources, which can be inferred to the number of stars in the pixel and the antipode pixel, respectively. Therefore, the pixels within the Galactic plane have high values of microlensing probability because there are a large number of stars on the Galactic plane. There are four hot spots outside the Galactic plane. They are the locations of the LMC, the SMC and their antipodes, which also have a huge number of observers or sources. However, the observers around the ecliptic plane have lower microlensing probability and discovery rate values as the Earth-Sun separation in the point of view of the ecliptic plane observer is smaller than the observer around the ecliptic poles. The smaller separations provide smaller caustic sizes, which lead to the lower probability and discovery rate values.

Around the Galactic plane, the average caustic crossing time is longer than the other areas. The long duration in the direction toward the Galactic plane can be explained by disc-disc observer-source pairs where the observer, the source and the Sun as the lens are moving in the same direction with similar speed. For the microlensing discovery rate, the map has a similar pattern to the microlensing probability map as the variation of average crossing time is much smaller than the variation in microlensing probability. 

\begin{figure}
\centering
\includegraphics[width=0.5\textwidth]{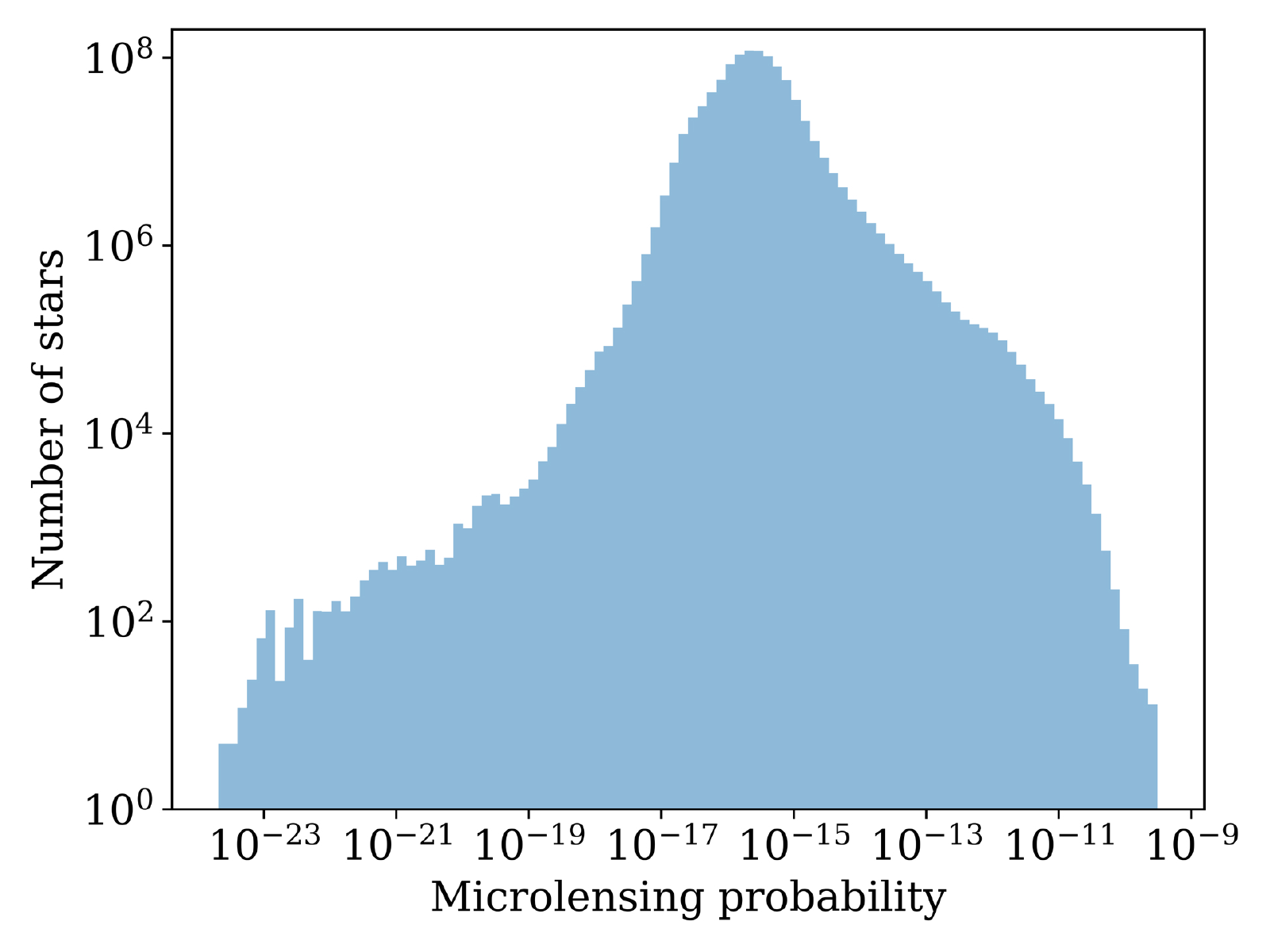} \\
\includegraphics[width=0.5\textwidth]{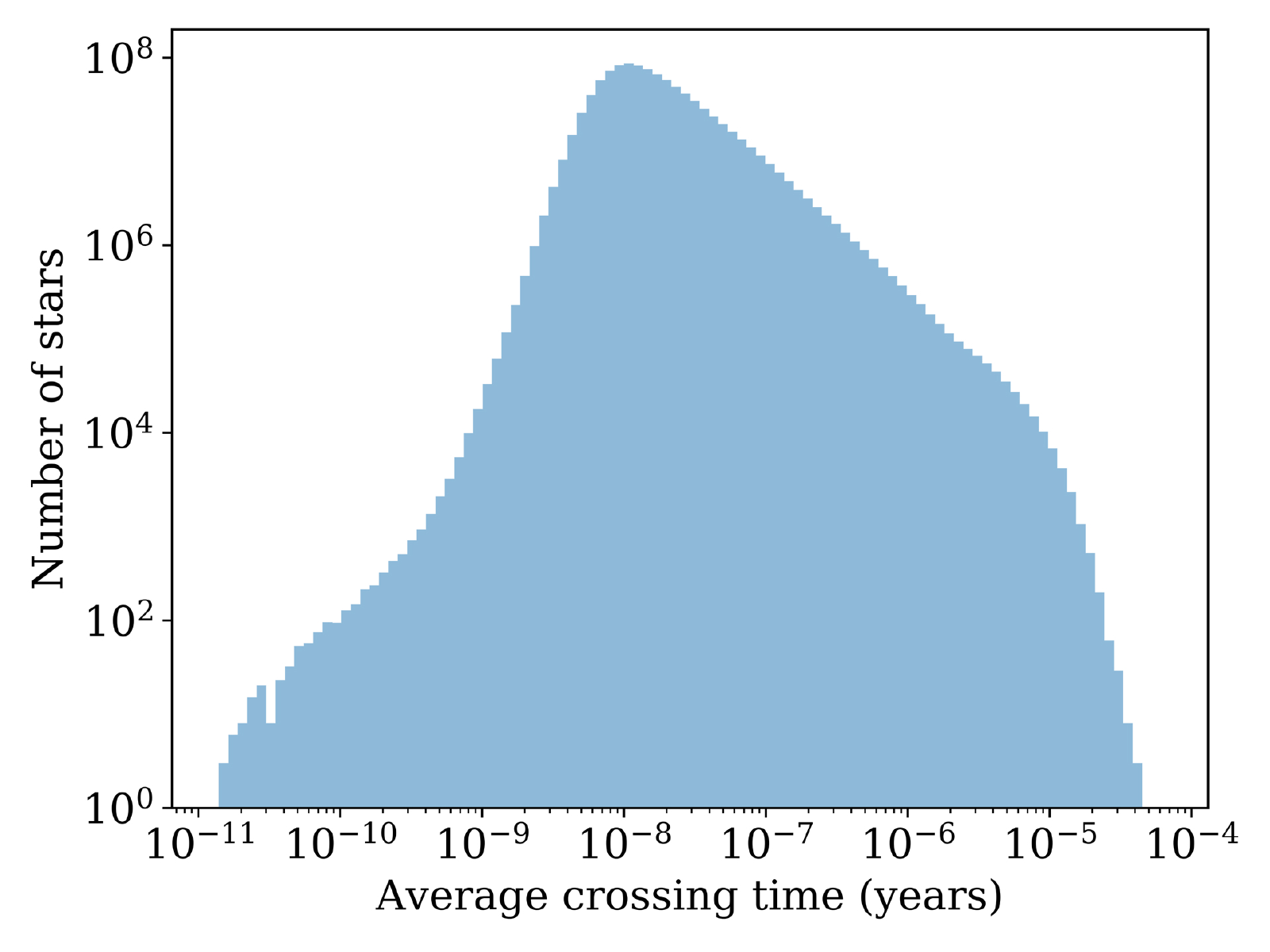} \\
\includegraphics[width=0.5\textwidth]{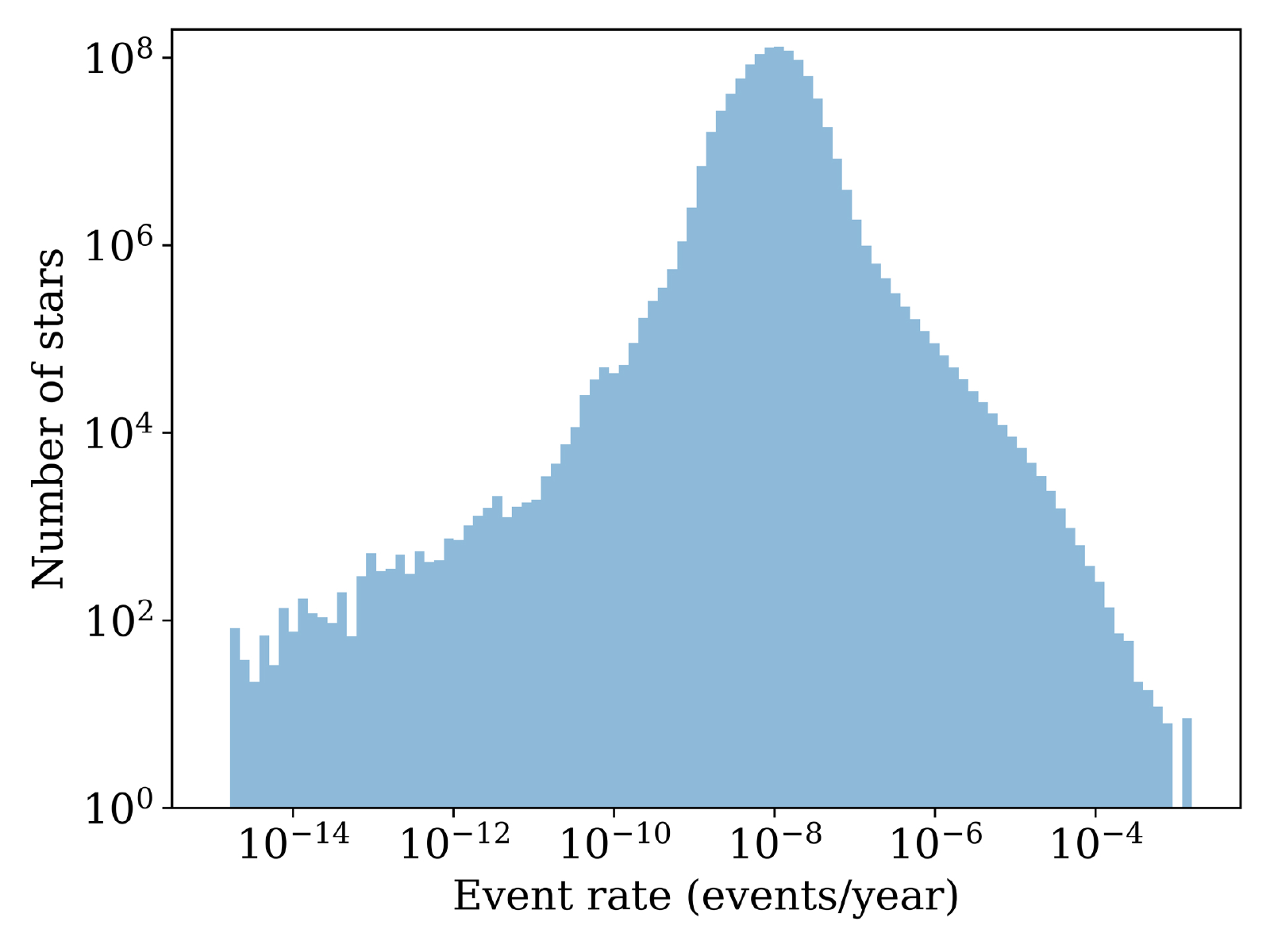}
\caption{Microlensing probability (Top), average caustic crossing time (Middle) and discovery rate (Bottom) distributions of all observers. The mode of microlensing probability, average caustic crossing time, discovery rate are $10^{-16}$, $10^{-8}$ yr and $10^{-8}$ events/yr, respectively. The total Earth discovery rate is $14.7$ observers per year.}
\label{Fig:DistPix}
\end{figure}

\begin{figure}
\centering
\includegraphics[width=0.5\textwidth]{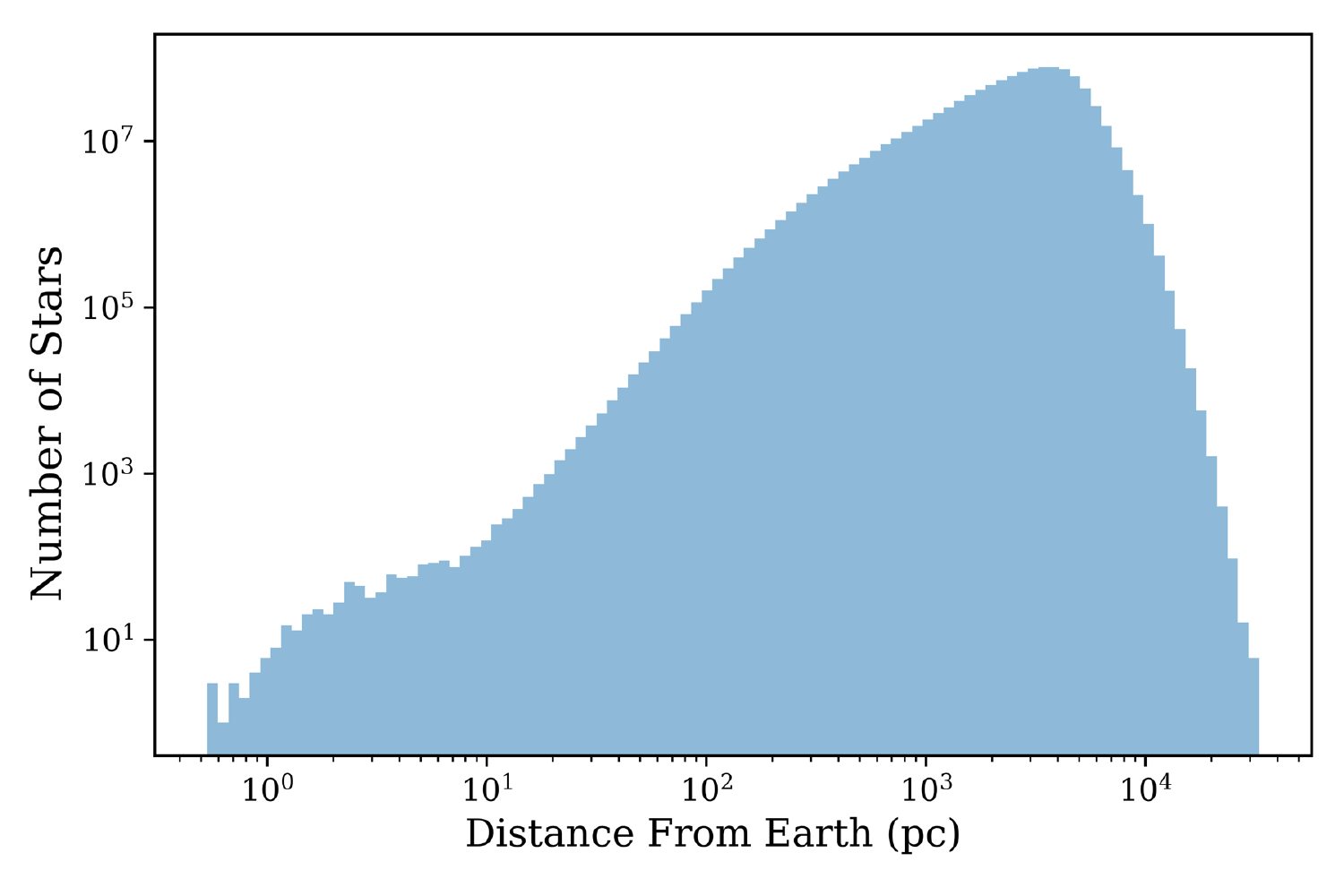}
\caption{Distribution of distance of stars from the Earth with magnitude limit of $G<20$ from \textit{Gaia}. The shape of the distance distribution is similar to the shape of microlensing properties in \Cref{Fig:DistPix}.}
\label{Fig:DistanceDist}
\end{figure}

In \Cref{Fig:DistPix}, the distributions of the microlensing parameters of all observers are presented. The microlensing probability covers the values between $2\times10^{-24}$ and $3\times10^{-10}$ with a peak at $10^{-16}$. The average crossing time ranges between $10^{-11}$ and $6\times10^{-5}$ yr with the mode at $10^{-8}$ yr, which is typical for Galactic planetary microlensing events. Combining the microlensing probability and average crossing time distributions, the discovery rates are between $10^{-15}$ and $10^{-3}$ observers/yr/deg$^{2}$. The shape of the distribution curves can be explained by the distance distribution with magnitude limit of $G<20$ (\Cref{Fig:DistanceDist}). 

For the whole sky, the total Earth discovery rate per year is 14.7 observers, which implies that on the average there is an extra-solar observer that can detect the Earth as a microlensing planet tens in a year. However, as \textit{Gaia} performs the observation in the $G$-band, the extinction effect can be seen in the low Galactic latitude. The observations in near-Infrared or Infrared bands should provide higher microlensing probabilities and discovery rates at low Galactic latitude due to less effects from the extinction on the observations. 

From 49 152 pixels, the pixel number 14 545, which is centred at RA = $296.01563^\circ$, Dec $34.22887^\circ$ ($l=69.00127^\circ$, $b=5.106634^\circ$), provides the highest microlensing probability and discovery rate of $3.28~\times~10^{-10}$ and $2.35~\times~10^{-2}$~observers/yr/deg$^{2}$, respectively. The microlensing probability value of the pixel is higher than the average, due to its location toward the Orion-Cygnus arm, but the average crossing time of the pixel is similar to the values of other pixels. As a result, the discovery rate distribution of this pixel is higher than others.

\subsection{Earth microlensing zones}

\begin{figure}
\centering
\includegraphics[width=0.5\textwidth]{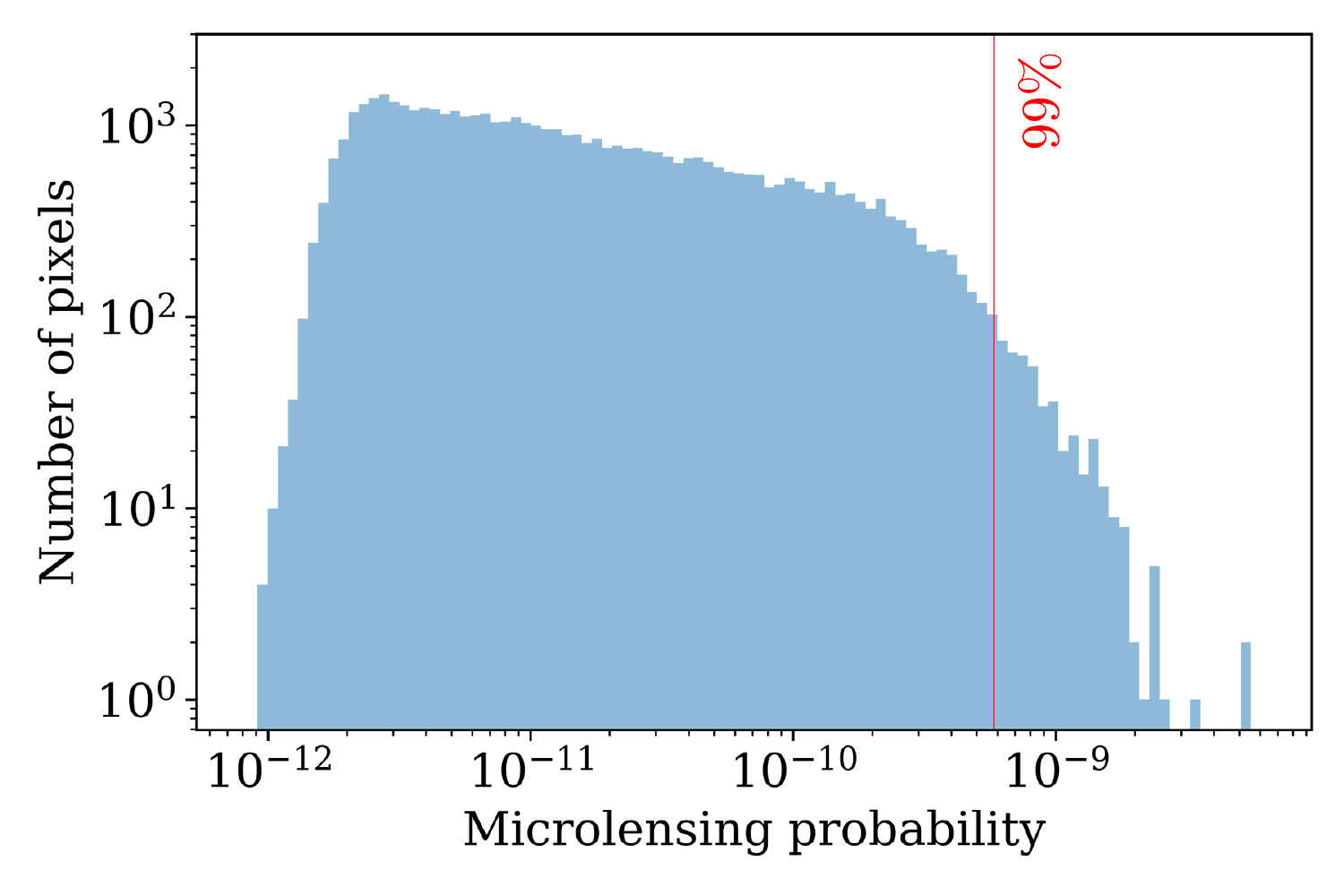} \\
\includegraphics[width=0.5\textwidth]{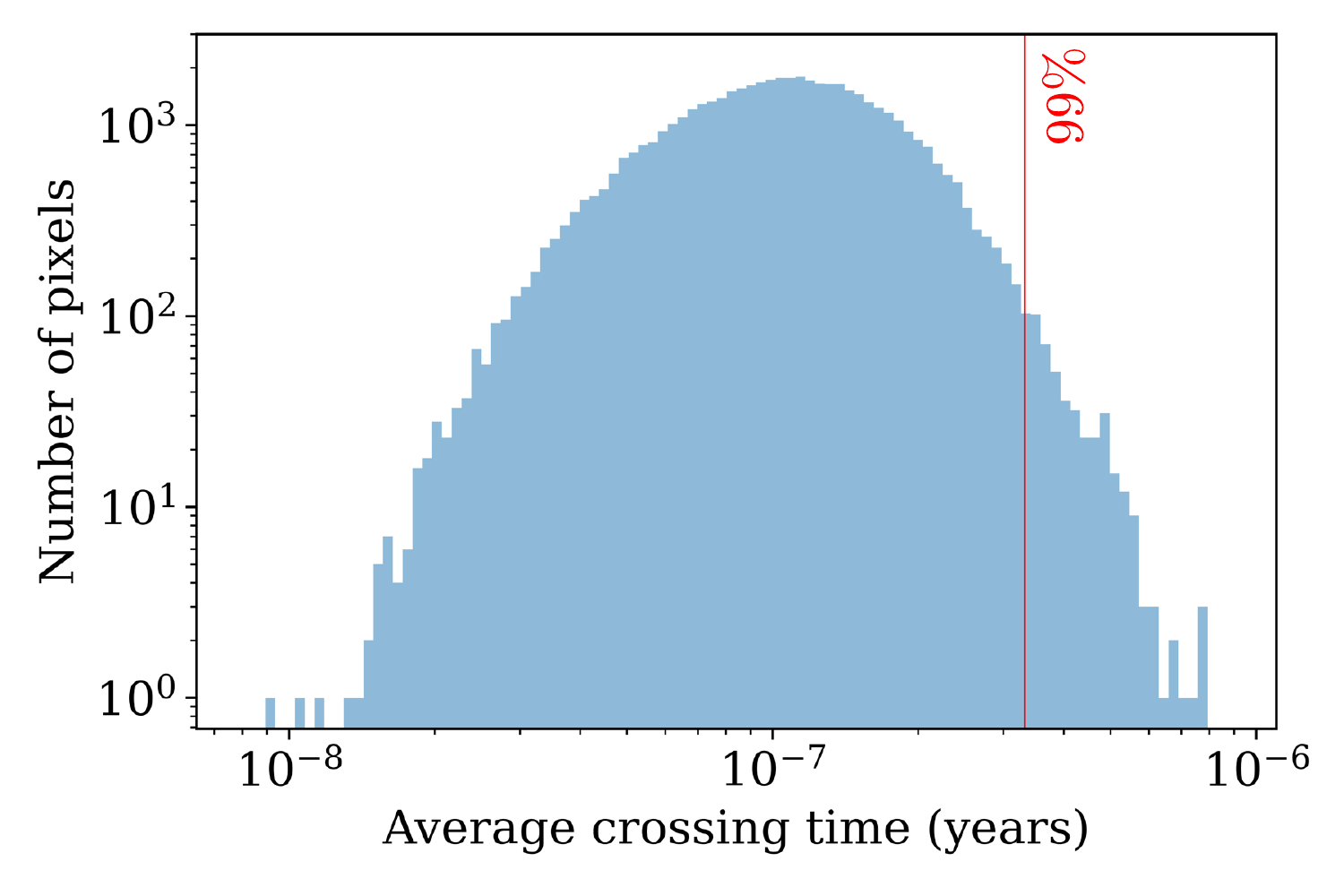} \\
\includegraphics[width=0.5\textwidth]{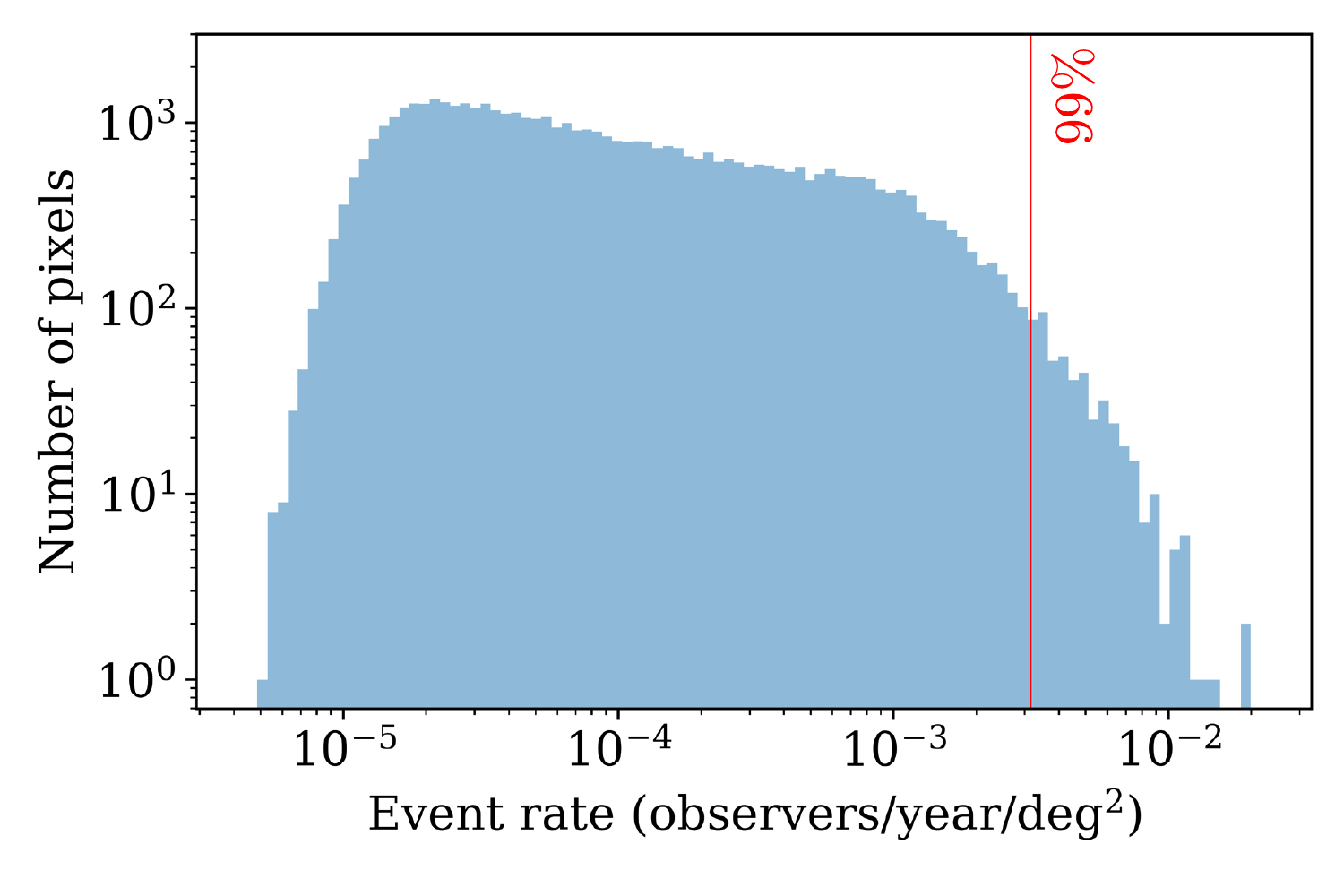}
\caption{Microlensing probability (Top), average caustic crossing time (Middle) and discovery rate (Bottom) distributions of HEALPix pixels. The red vertical line shows the 99$^\textup{th}$ percentile thresholds (492 pixels of HEALPix scheme at level 6) of each distribution. The 99$^\textup{th}$ percentile thresholds of the discovery rate indicates the threshold of the EMZs.}
\label{Fig:99Percent}
\end{figure}

Recall that the region in which the Earth could be detected from the extraterrestrial observers with microlensing technique, called the ``Earth Microlensing Zone'', is defined as the area which contains the top 1\% HEALPix pixels ordered by microlensing discovery rates. The Earth microlensing zones (EMZs) cover approximately 413 deg$^2$ of the entire sky or 492 pixels of the HEALPix scheme at level 6. In \Cref{Fig:99Percent}, the distributions of microlensing probability, average crossing time and discovery rates are shown with the threshold of the 99$^\textup{th}$ percentile. The thresholds are at $5.82~\times~10^{-10}$, $3.32~\times~10^{-7}$~yr and $3.16~\times~10^{-3}$~observers/yr/deg$^2$ for microlensing probability, average crossing time and discovery rates, respectively. Although the EMZs cover only 1\% of the sky area, the EMZs contain the Earth discovery rate of 2.42 observers per year, which corresponds to 16\% of the discovery rate of the whole sky. 

\begin{figure*}
\centering
\includegraphics[width=0.9\textwidth]{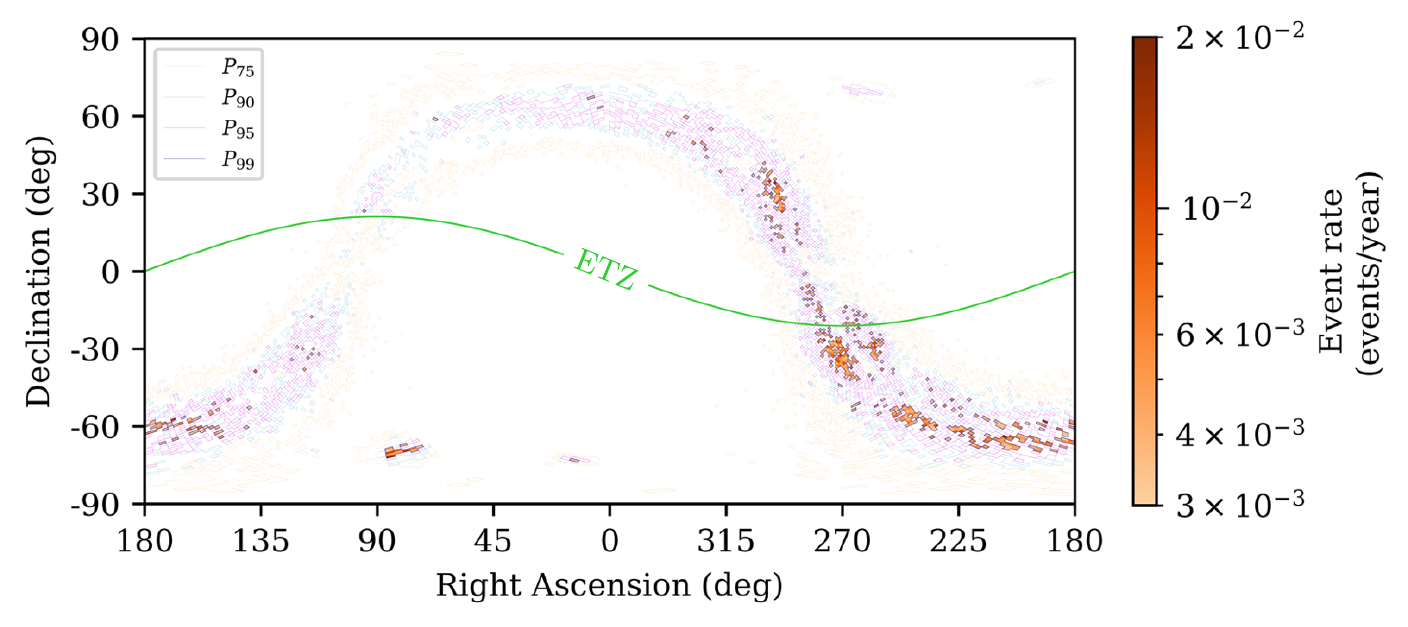}
\caption{The EMZs, which are defined by the areas which contain the top 1\% HEALPix, are filled with colour gradients showing corresponding microlensing discovery rates. The light orange, light blue, pink, and navy contours represent the discovery rate distributions at the 75th, 90th, 95th and 99th percentiles, respectively. The lime green line shows the ETZ. The EMZs are located in the Galactic plane, the LMC, and the SMC.}
\label{Fig:EMZ}
\end{figure*}

The map of the EMZs is shown in \Cref{Fig:EMZ} together with the Earth Transit Zone (ETZ) defined by \cite{hel2016}. The EMZs are located at low Galactic latitude area ($|b| < 10^\circ$), the LMC and the SMC. The majority of EMZs are located in the direction of Galactic centre. A SETI survey has already been performed in this region under the Breakthrough Listen program \citep{gaj2021} along with monitoring from a number of microlensing surveys (MOA \citep{bon2001,sum2003}, OGLE \citep{uda1993}, KMTNet \citep{lee2015} and K2-Cycle9 \citep{mcd2021}). In the Galactic plane, the EMZs locate not only at the direction of the Galactic bulge, but also in the directions of Carina-Sagittarius arm and the Scutum-Centaurus arm, which Earth microlensing events occur in front of dense sources residing in the Perseus arm. However, there is a $\sim1^\circ$ gap between the northern and southern Galactic hemispheres. The gap might be caused by the effects of interstellar extinction which is high in the \textit{Gaia} optical $G$-band.

In addition to the EMZs which are indicated by the 99$^\textup{th}$ percentile contours, the 95$^\textup{th}$, 90$^\textup{th}$ and 75$^\textup{th}$ percentile contours are shown in \Cref{Fig:EMZ} to present the areas that have high Earth discovery rate. In the map, there are areas with the discovery rate distribution above the 90$\textup{th}$ percentile, including a part of the EMZs, that overlap with the proposed ETZ near the Galactic centre, which could be the area of interest for future SETI search. Moreover, the upcoming microlensing surveys, such as \textit{Roman} and \textit{Euclid} \citep{pen2013,pen2019,bac2022} might detect the signals from Earth-like planets which could be the future SETI targets in this area.

\begin{figure*}
\centering
\includegraphics[width=0.9\textwidth]{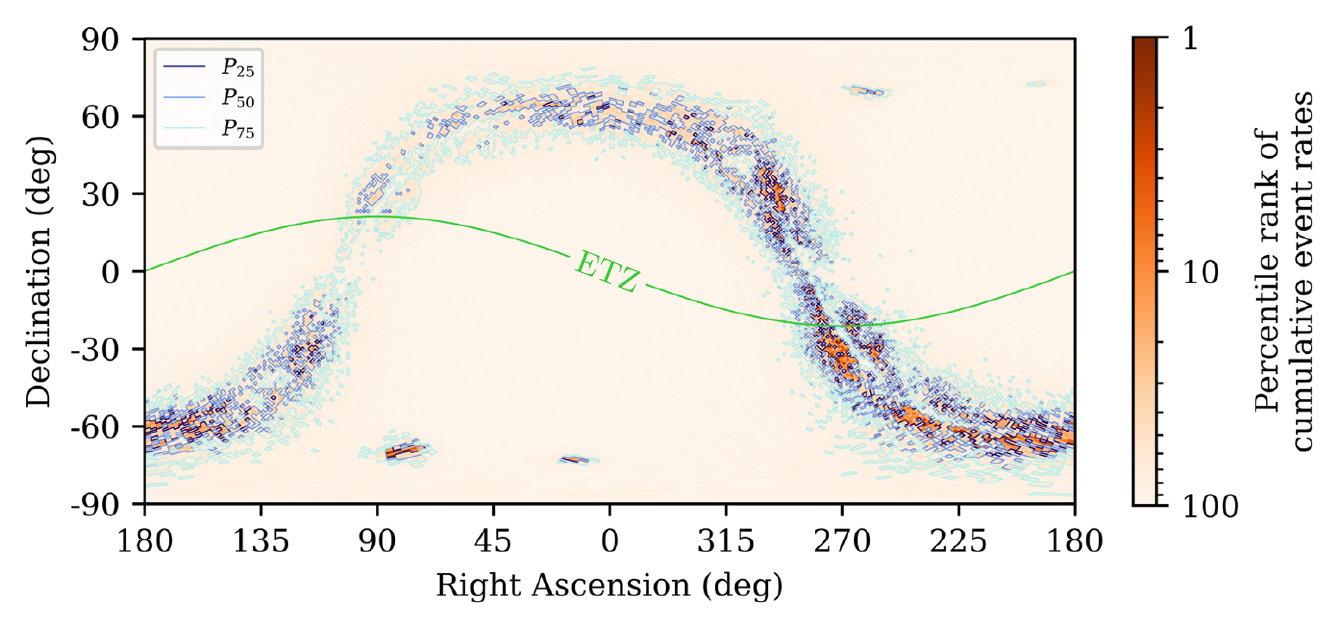}
\caption{Percentile rank of cumulative discovery rates with the ETZ (light green line). The turquoise, light blue, and navy contours represent the 25th, 50th and 75th cumulative percentiles, respectively. As 75th cumulative percentiles contours locate in the Galactic disc, Galactic bulge, the LMC and the SMC, the Earth is likely to be detected by the observes in these areas.}
\label{Fig:PercentileRank}
\end{figure*}

In \Cref{Fig:PercentileRank}, the percentile rank of cumulative discovery rates is shown in order to present the region of hotspots with high Earth discovery rates for the entire sky. The percentiles are calculated from the cumulative sums in descending order. More than 75\% of the discovery rates are located in the Galactic plane, the LMC, the SMC, and their antipodes. Therefore, this can be confirmed that the Earth is likely to be detected by the observers with photometric microlensing in the Galactic disc, Galactic bulge, the LMC, the SMC, and their antipodes.

From billions of stars, the stars which are located in the EMZs with highest microlensing discovery rates are listed in \Cref{Tab:Top10}. Although ten stars with highest discovery rates are listed, the interested stars to search for extraterrestrial intelligence are not limited to these stars. All stars within the Galactic plane can be interested in the SETI search. However, in order to find habitable planets, the planets should be in the habitability zone of the Galaxy, called the ``Galactic Habitable Zone''. The Galactic Habitable Zone (GHZ) has been proposed to be between 7 kpc and 9 kpc from the Galactic centre \citep{lin2004}. Observers with highest discovery rates are located considerably close to the Earth. All of the stars in the list are located closer than 20 pc within the same Galactic environments as the Solar system.

\begin{figure*}
\centering
\includegraphics[width=0.49\textwidth]{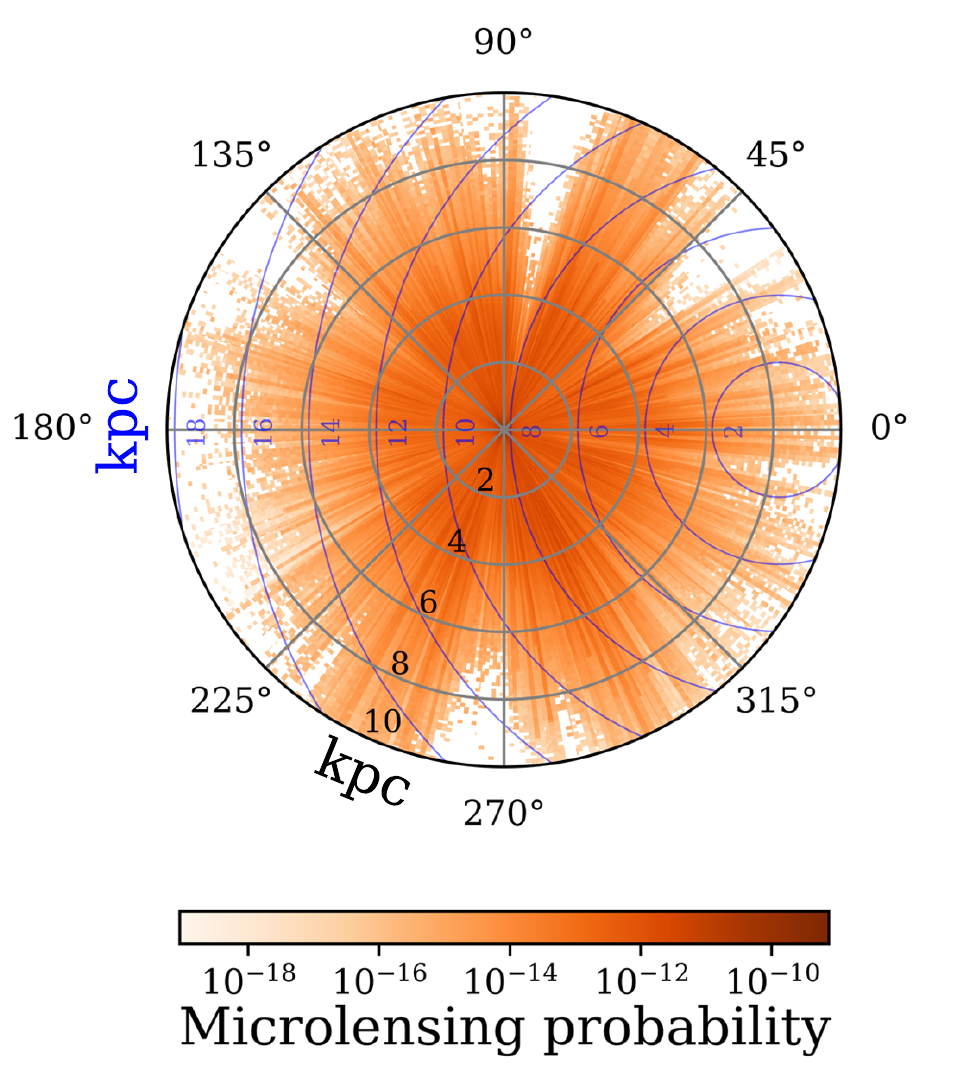}
\includegraphics[width=0.49\textwidth]{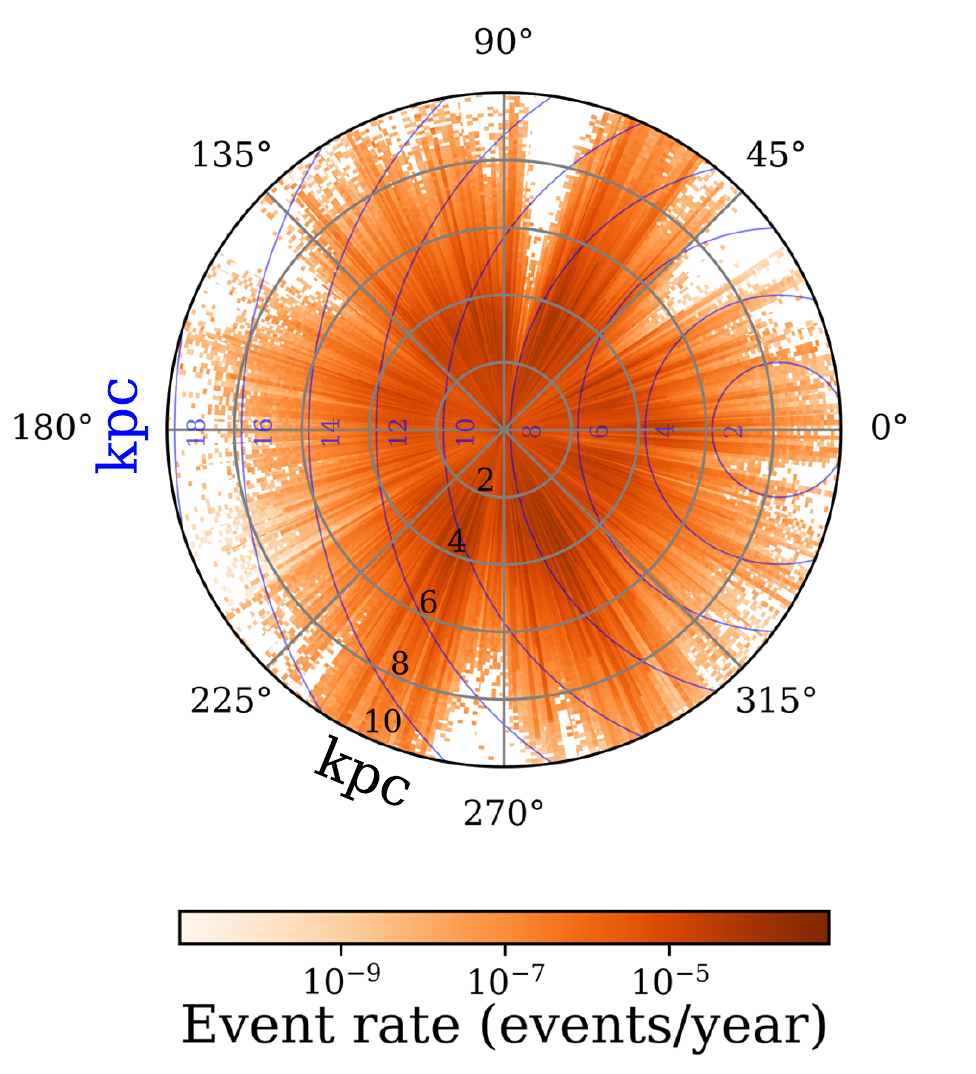} \\
\includegraphics[width=0.49\textwidth]{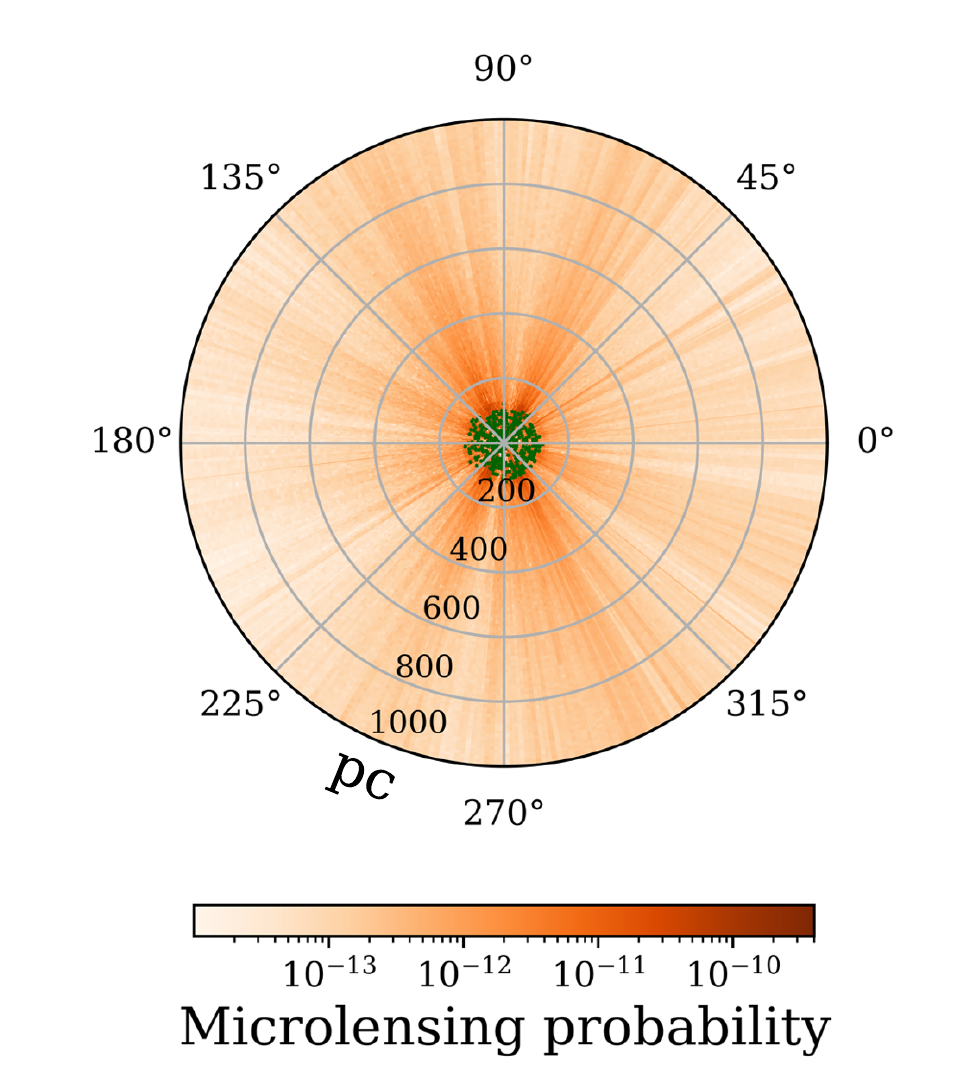}
\includegraphics[width=0.49\textwidth]{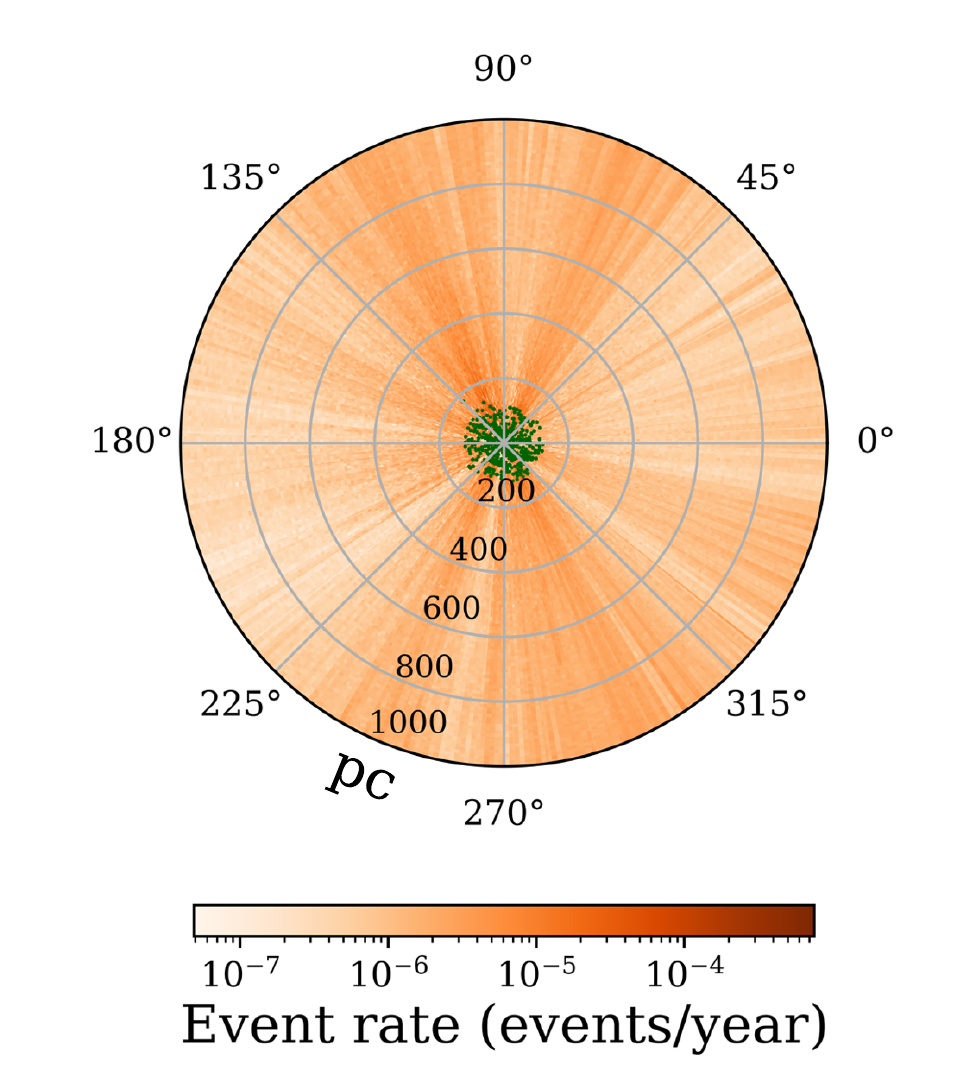}
\caption{Microlensing probability (Left) and discovery rate (Right) as a function of the distance from the Earth in different directions along the Galactic plane. The figures show the top view of the Milky Way galaxy with the Galactic longitude system centered on the Earth and the Sun. The size along the azimuthal axis is the angular size of HEALPix pixels in the Galactic plane. Top figures show the microlensing properties averaged among observers located within distance bins of 100 pc each, up to 10 kpc. The blue lines designate the distance from the Galactic centre in kpc. Bottom figures are similar to the top figures, but the radial bins are now of size 10 pc each, up to 1 kpc. Radial bins with the maximum averaged probability and rate in each HEALPix are marked with dark green. The observers with the highest values of microlensing probability and discovery rate are mostly located within the distance of 100~pc away from the Earth, which is within the range the strongest sources that isotropically radiate radio-waves can be detected with the SKA.}
\label{Fig:Polar}
\end{figure*}

The microlensing probabilities and discovery rates as functions of distance from the Earth on the Galactic plane are shown in \Cref{Fig:Polar}. The figures show the parameters of observers located within every 100~pc and 10~pc intervals from the Earth up to 10~kpc and 1~kpc, respectively. Nearby observers within 200~pc from the Earth possess higher microlensing probabilities and discovery rates than further observers. These observers are located in the same Galactic environments as the Solar system. The observers with the highest values of microlensing probability and discovery rate are mostly located within the distance of 100~pc away from the Earth. The peaks suggest that the observers who can detect the Earth as a microlensing planet are not located too far to detect their radio signals. Using military radars with maximum power per transmitter ($P_{\textup{max}} \simeq 2\times10^{11}$ W) which might be the strongest sources that radiate radio-waves isotropically, the maximum distance that the signals can be detected with the SKA is $\sim$1.26 kpc \citep{rah2016}. Therefore, the signals from extraterrestrial intelligent observers are strong enough to be detected from the SETI search.

\begin{table*}
    \centering
    \caption{Stars with the top 10 highest microlensing discovery rates retrieved from calculations done at HEALPix scheme levels 6.}
    \begin{tabular}{ c l r r c c c}
         \hline
         Rank & \textit{Gaia} \texttt{source\_id} & \multicolumn{1}{c}{RA ($^\circ$)}& Dec ($^\circ$) & \begin{tabular}{@{}c@{}}Microlensing probability\\ $(\times 10^{-10})$ \end{tabular} & \begin{tabular}{@{}c@{}} Earth discovery rate \\ ($\times 10^{-3}$ observers/yr/deg$^{2}$) \end{tabular}  & Distance (pc) \\
         \hline
         1  & 470826482635704064  & 67.80884  & 58.96826  & 2.45 & 1.60 & 5.54 \\
         2  & 470826482635701376  & 67.81353  & 58.96975  & 2.46 & 1.56 & 5.52 \\
         3  & 527956488339113472  & 8.14240   & 67.23438  & 2.35 & 1.33 & 9.86 \\
         4  & 2009481748875806976 & 348.33720 & 57.16963  & 2.46 & 1.32 & 6.53 \\
         5  & 5853498713160606720 & 217.39347 & -62.67618 & 2.75 & 1.29 & 1.30 \\
         6  & 1872046574983507456 & 316.74774 & 38.76341  & 1.28 & 1.28 & 3.50 \\
         7  & 527956488339113600  & 8.14223   & 67.23331  & 2.33 & 1.27 & 9.96 \\
         8  & 1872046574983497216 & 316.75293 & 38.75563  & 1.28 & 1.26 & 3.49 \\
         9  & 1638979384378696704 & 267.01666 & 70.88142  & 3.05 & 1.23 & 6.21 \\
         10 & 523433578540463488  & 16.79932  & 63.94270  & 1.71 & 1.08 & 15.1 \\
         \hline
    \end{tabular}
    \label{Tab:Top10}
\end{table*}

\subsection{Known exoplanets in the Earth microlensing zone}

\begin{figure*}
\centering
\includegraphics[width=0.9\textwidth]{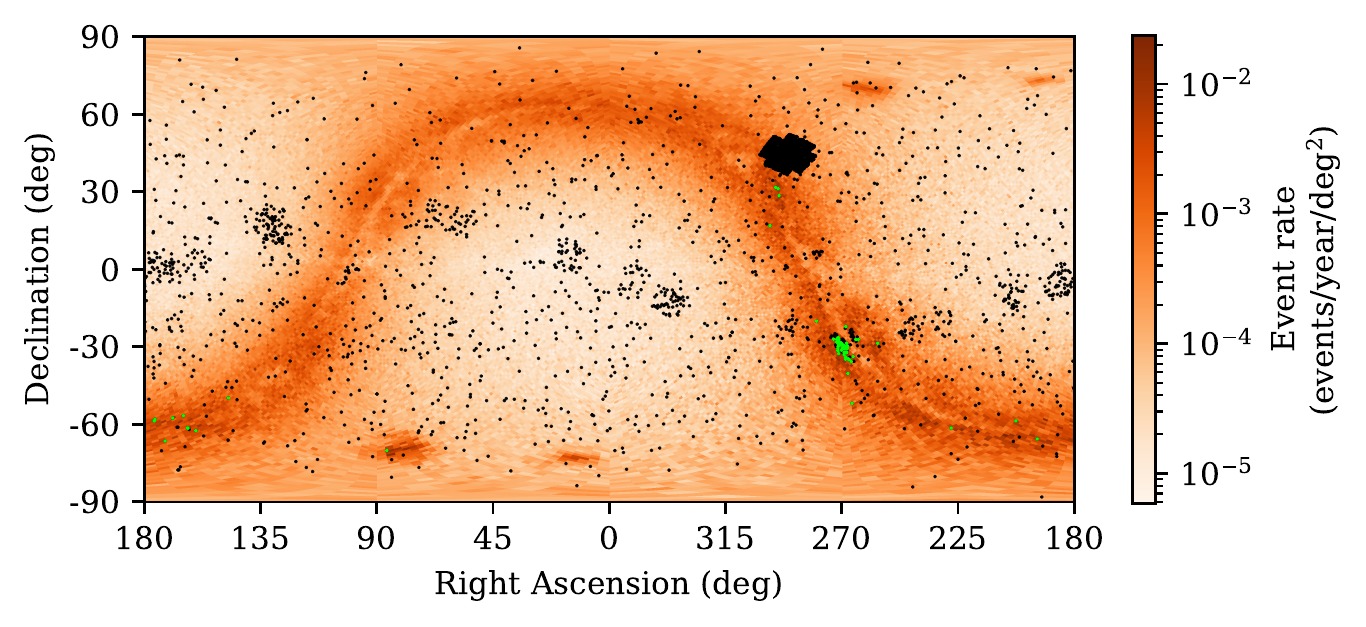}
\caption{Earth discovery rate map marked with confirmed exoplanets (black dots). Exoplanets that are located in the Earth Microlensing Zone are marked in green. In total, 85 confirmed exoplanets are located within the EMZs.}
\label{Fig:Known_Planets}
\end{figure*}

To date, around 5000 exoplanets have been confirmed orbiting 3500 stellar hosts. The known planets are over-plotted on the Earth discovery rate map calculated using HEALPix scheme at level 6 (\Cref{Fig:Known_Planets})\footnote{The exoplanets and their hosts data was retrieved from \texttt{https://exoplanetarchive.ipac.caltech.edu/} on May 23, 2022.}. Eighty five confirmed exoplanets are located in the EMZs. The information of the planets are shown in \Cref{Tab:Known_Planets}. From the list, all planetary systems are located in the Galactic plane, where the high discovery rates are found, and also locate within the distance of 1.5 kpc from us, which are in the same Galactic environments as the Solar system.

MOA-2007-BLG-192L b is a known exoplanet located within the HEALPix number 28 869, which provides the highest discovery rate among other HEALPix pixels with known exoplanets. It is a super-Earth approximately 3.2 times Earth's mass, orbiting a very low mass late-type M-dwarf, approximately 660 pc away in the direction of the bulge \citep{kubas2012}. With a planetary radius of 1.63 Earth radius, the planet might be a rocky planet which might be suitable for hosting life. However, the planet has a star-planet separation of 0.66 AU which is beyond the snow line of the M-dwarf host star.

GJ 422 b is in the HEALPix ranked the second in the list. The planet is a Neptune-like exoplanet with a minimum mass of 9.9 Earth-mass that orbits an M-type star, in the direction of the Perseus arm. The planet has an orbital distance of 0.119 AU from the host star which is in the habitable zone of the system. Although the planet is a Jovian planet, rocky moons orbiting around it can host life \citep{chy1997,wil1997}. With the distance of 13 pc from the Earth, the radio signature from GJ 422 b can be detected easily on the Earth. Rocky planets or moons that have high potential to host life forms which are of interest for the SETI search. In the future, exoplanetary detection surveys in the direction of EMZs, especially in the direction of the Galactic bulge (e.g. \textit{Roman} and \textit{Euclid} \citep{pen2013,pen2019,bac2022}), will increase the number of rocky planets or exomoons that are the SETI targets in these areas.

\begin{table*}
    \centering
    \caption{Confirmed exoplanets in the Earth Microlensing Zone, ranked with respect to the Earth discovery rates of their HEALPix pixels for which they are in. The table also shows the corresponding microlensing probabilities of their HEALPix pixels.}
    \begin{tabular}{c l @{\hspace{0cm}} c @{\hspace{0.2cm}} c @{\hspace{0.2cm}} c @{\hspace{0.2cm}} c @{\hspace{0.2cm}} c @{\hspace{0.2cm}} c @{\hspace{0.2cm}} l}
    \hline
         \multirow{3}{*}{Rank} & \multirow{3}{*}{System name} & \multirow{3}{*}{HEALPix} & \multirow{3}{*}{\shortstack{Microlensing\\ probability\\ $(\times 10^{-9})$  }} & \multirow{3}{*}{\shortstack{Earth discovery rate\\($\times 10^{-2}$)\\(observers/yr/deg$^{2})$}} & \multirow{3}{*}{\shortstack{Host mass\\ (M$_\odot$)}} & \multirow{3}{*}{\shortstack{Planet mass \\ (M$_\textup{J}$)}}  &\multirow{3}{*}{\shortstack{Distance\\ (pc)}}  & \multicolumn{1}{c}{\multirow{3}{*}{References}} \\
         & & & & & & & \\  
         & & & & & & & \\
         \hline
         1 & MOA-2007-BLG-192L b     & 28 869 & 2.16 & 1.51 & 0.084 & 0.010 & 660 & \cite{bennett2008}, \\
           &  &  &  &  &  &  &  & \cite{kubas2012}\\
         2 & GJ 422 b                & 37 942 & 0.538 & 1.42 & 0.35 & 0.035 & 13 & \cite{tuomi2014}, \\
           &  &  &  &  &  &  &  & \cite{feng2020}\\
         3 & OGLE-2015-BLG-1649L b   & 28 726 & 0.986  & 1.38 & 0.34 & 2.5 & 4200 & \cite{nagakane2019} \\
         4 & HD 39194 b        & 33 091 & 0.497  & 1.32 & 0.67 & 0.013 & 26 & \cite{unger21} \\
           & HD 39194 c        & & & & 0.67 & 0.020 & 26 & \cite{unger21} \\
           & HD 39194 d        & & & & 0.67 & 0.013 & 26 & \cite{unger21} \\
         5 & MOA-2016-BLG-227L b     & 28 868 & 2.28  & 1.05 & 0.29 & 2.8 & 6500 & \cite{koshimoto2017}\\
           & OGLE-2012-BLG-0563L b   &       &       &      & 0.34 & 0.39 & 1300 & \cite{fukui2015}\\
         6 & MOA-2010-BLG-353L b     & 28 870 & 1.69  & 1.04 & 0.18 & 0.27 & 6400 & \cite{rattenbury2015}\\
           & MOA-2011-BLG-322L b     &       &       &      & 0.39 & 12 & 7600 & \cite{shvartzvald2014}\\
         7 & OGLE-TR-111 b           & 37 247 & 0.399 & 0.974 & 0.85 & 0.55 & 1100 & \cite{pont2004},\\
            &  &  &  &  &  &  &  & \cite{bonomo2017}\\
           & OGLE-TR-113 b           &       &       &      & 0.78 & 1.3 & 570 & \cite{bouchy2004},\\
            &  &  &  &  &  &  &  & \cite{bonomo2017}\\
         8 & HD 331093 b             & 14 447 & 0.380  & 0.853 & 1.0 & 1.5 & 50 & \cite{dalal21} \\
         9 & HD 165155 b             & 28 779 & 1.91  & 0.84 & 1.0 & 2.9 & 65 & \cite{jenkins2017}\\
           & MOA-2011-BLG-028L b     &       &       &      & 0.80 & 0.094 & 7300 & \cite{skowron2016}\\
           & MOA-2013-BLG-220L b     &       &       &      & 0.88 & 2.7 & 6700 & \cite{yee2014}, \\
           &  &  &  &  &  &  &  & \cite{vandorou2020}\\
           & OGLE-2003-BLG-235L b    &       &       &      & 0.63 & 2.6 & 5800 & \cite{bond2004}, \\
           &  &  &  &  &  &  &  & \cite{bennett2006}\\
         10 & KMT-2017-BLG-1146L b   & 28 729 & 0.944 & 0.826 & 0.40 & 0.85 & 6600 & \cite{shin19}\\
    \hline
    \end{tabular}
    \label{Tab:Known_Planets}
\end{table*}

\subsection{Microlensing properties of an observer}

The microlensing probabilities and Earth discovery rates presented in \Cref{sec:properties_map} are calculated from the total microlensing probability and discovery rate in each HEALPix pixel. The values represent the microlensing probabilities and discovery rates of all stars in the area, which depend on the number of observer stars. The areas with high density of stars provide higher values of the microlensing probabilities and discovery rates. In order to obtain the properties of an individual star in the HEALPix, the median values of the properties in each area can be an appropriate representative of the individual microlensing probability and the discovery rate in that area,

\begin{equation}
\tilde{P} = \textup{Median}(P_\textup{o}) \ ,
\end{equation}
where $P$ can be a microlensing probability or discovery rate.

\begin{figure*}
\centering
\includegraphics[width=0.45\textwidth]{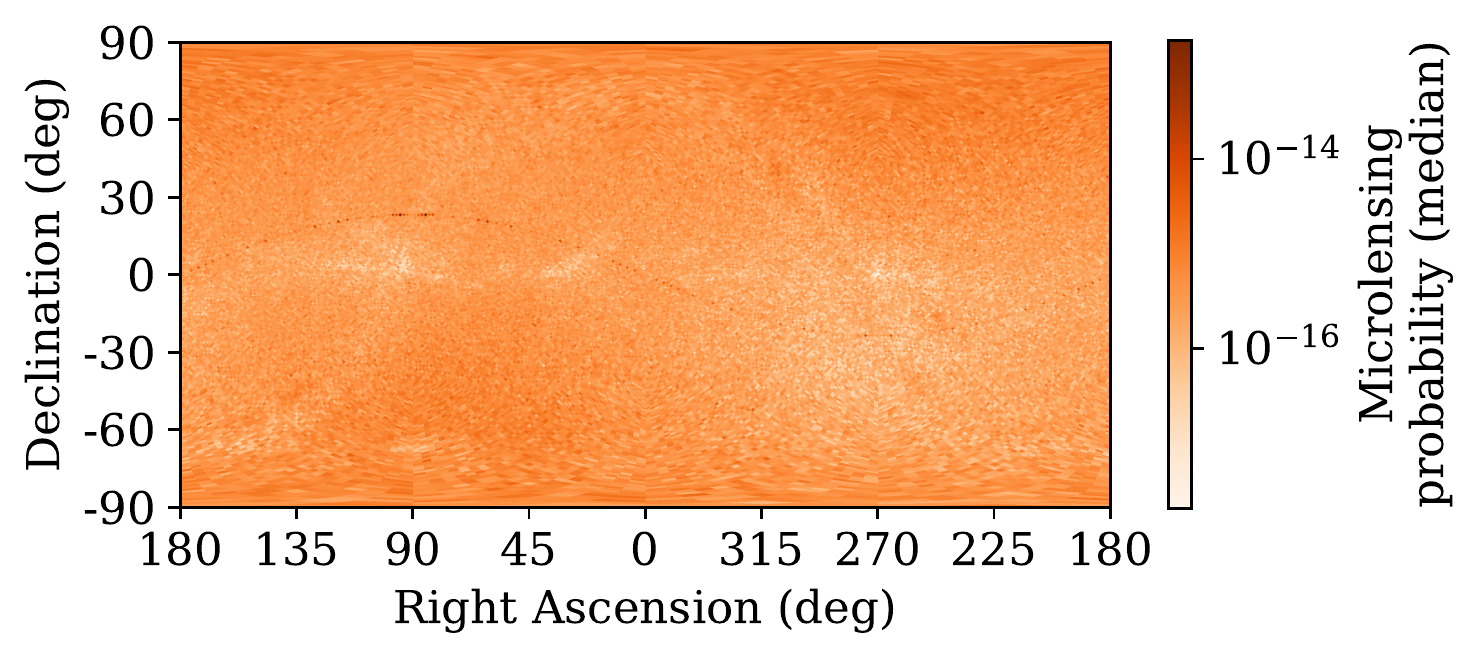}
\includegraphics[width=0.45\textwidth]{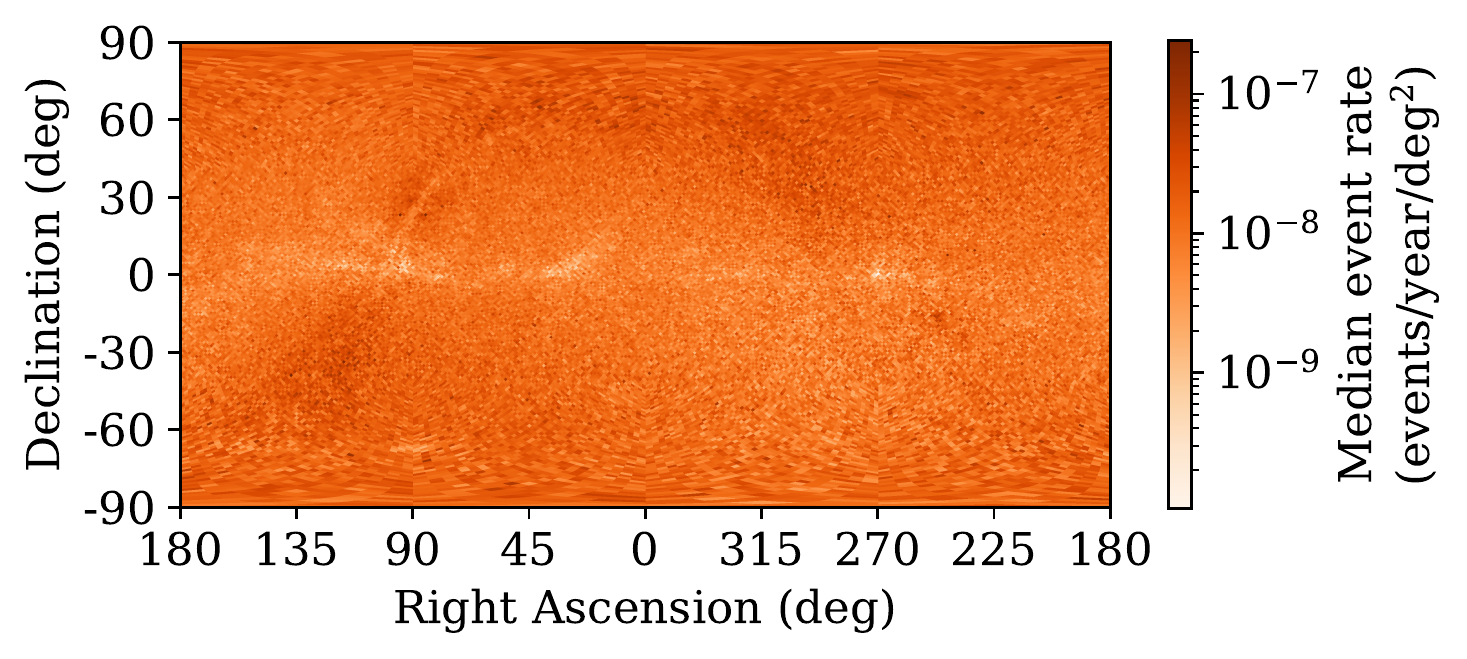}
\caption{Median microlensing probability (Left) and discovery rate (Right) maps from calculations corresponding to HEALPix scheme level 6. The maps show that extraterrestrial observers located at the antipodes of the Galactic centre, the LMC and the SMC have a higher rate to detect the Earth as a microlensing planet than the other areas.}
\label{Fig:Median}
\end{figure*}

The median microlensing probability and discovery rate maps are shown in \Cref{Fig:Median}. Comparing these with the total microlensing probability and discovery rate maps in \Cref{Fig:MapHP6HP12}, the Galactic plane and the antipodes of the LMC and the SMC still provide high microlensing probabilities and discovery rates. Unlike the total probability and rate maps, the Galactic centre, the LMC, and the SMC do not provide high median values. In this region, a larger number of observers produces high total probability and rate, but probability and rate for the individual observer are not high due to a fewer number of sources. On the other hand, the number of observers is small in the antipodes whereas the number of sources is high. Therefore, both total and median probability and rate of each observer are high. This can be implied that extraterrestrial intelligences located at the antipodes of the Galactic centre, the LMC, and the SMC have a higher rate to detect the Earth as a microlensing planet than the other areas.

\section{Conclusions}
\label{sec:conclusion}

With our current technology, microlensing technique is the best method to detect distant planets located around the most common stars in our Galaxy. In theory, a long range detection method like microlensing could be employed by technological civilisations to detect the Earth across Galactic distance scales. Locations from which Earth may be detectable are considered to be good candidates for targeted SETI surveys, in which case Earth's microlensing signature could make even distant stars worth considering for such surveys. In this work, the ``Earth Microlensing Zone'' (EMZ), the areas of our sky from which Earth's microlensing signal might most likely be detected by a distant observer, are evaluated. Similar to the ``Earth Transit Zone'' (ETZ), the EMZ might offer a potential target area for SETI projects.

We use a catalogue of more than a billion stars from the \textit{Gaia} DR2 to calculate the Earth microlensing detection probability to other observers and rates at which observers may discover us, assuming that all stars are equally likely to host a technological civilization. The HEALPix scheme is applied to divide the sky into small areas. We construct HEALPix maps of different resolutions and find that they give similar microlensing discovery rates. The microlensing probability and discovery rate maps show that both properties have high values around the Galactic plane, the large magellanic cloud, the small magellanic cloud and at their antipodes. The total Earth discovery rate is shown to be 14.7 observers per year over the entire sky, assuming an observer around every \textit{Gaia} stars with $G<20$. This means that, on average, a microlensing signal due to the Earth that is comparable in size to those we detect from other planets, occurs tens per year towards any star in the Galaxy. The direction with the highest microlensing probability and discovery rate is centred in the direction toward the Orion-Cygnus arm in the Galactic plane ($l=69.0^\circ$, $b=-5.1^\circ$) with the Earth microlensing probability and discovery rate values of $3.28\times 10^{-10}$ and $2.35\times10^{-2}$ observers/yr/deg$^{2}$, respectively. Overall, it seems that the Earth is very dark to photometric microlensing discovery by other observers, unless they have sensitivity well beyond our own present capabilities.

Defining the EMZs to be the areas that contain the top 1\% of the discovery rate, we find the optimal regions to be located at low Galactic latitude ($|b| < 10^\circ$) near the Galactic centre, the LMC and the SMC. The total microlensing discovery rate within the EMZ regions is $3.16\times10^{-3}$ observers/yr. The Galactic centre is performed the SETI survey under the Breakthrough Listen program \citep{gaj2021} and monitored by a number of microlensing surveys (MOA \citep{bon2001,sum2003}, OGLE \citep{uda1993}, KMTNet \citep{lee2015} and K2-Cycle9 \citep{mcd2021}. Moreover, in the future, the bulge will be monitored by large microlensing surveys, such as \textit{Roman} and \textit{Euclid} \citep{pen2013,pen2019,bac2022}, which will increase the number of detected Earth-like planets in the area. There are areas near the Galactic centre where the EMZs overlap with the previously proposed ETZ. The area is worth considering for targeted SETI surveys based on the Mutual Detectability approach \citep{ker2021}, as we might detect signals from extraterrestrial intelligent observers in the area.

Moreover, astrometric microlensing, which offers a much larger detection cross section than photometric microlensing, might increase the chance to detect the Earth for extraterrestrial observers. However, a proper calculation of this requires a detailed consideration of the combined astrometric microlensing signature of all Solar system planets in order to isolate the true long-range Earth microlensing signature to distant observers.
 
\section*{Acknowledgements}

This work presents results from the European Space Agency (ESA) space mission \textit{Gaia}. \textit{Gaia} data are being processed by the \textit{Gaia} Data Processing and Analysis Consortium (DPAC). Funding for the DPAC is provided by national institutions, in particular the institutions participating in the \textit{Gaia} MultiLateral Agreement (MLA). The \textit{Gaia} mission website is \texttt{https://www.cosmos.esa.int/gaia}. The \textit{Gaia} archive website is \texttt{https://archives.esac.esa.int/gaia}.

This work was supported by a National Astronomical Research Institute of Thailand (NARIT) and Thailand Science Research and Innovation (TSRI) research grant. This research was supported by Chiang Mai University. SS acknowledges the support of the Integrated Science Program of Hokkaido University. TC is supported by Science Classrooms in University-Affiliated School, Thailand. The authors acknowledge the anonymous referee for the valuable suggestions that helped to improve the paper.

\section*{Data availability}
The \textit{Gaia} DR2 microlensing probability, average crossing time and Earth discovery rate data are available in the article and in the online supplementary material.

%\nocite{*}

%%%%%%%%%%%%%%%%%%%% REFERENCES %%%%%%%%%%%%%%%%%%

\bibliographystyle{mnras}
\bibliography{EarthMicrolensingZone}

\appendix
\section{Calculated microlensing properties}
\begin{table*}
    \caption{Microlensing probability, average caustic crossing time and Earth discovery rate with HEALPix scheme at level 6 and level 12. The parameters are mapped on HEALPix level 6 index.}
    \label{Tab:MAPHP6HP12}
    \begin{tabular}{@{\extracolsep{4pt}} l l l l l l l l l }
    \hline
    \multirow{3}{1.2cm}{\centering Level 6 HEALPix index} & \multirow{3}{*}{\begin{tabular}{@{}l@{}} RA \\(deg)  \end{tabular}} & \multirow{3}{*}{\begin{tabular}{@{}l@{}} Dec \\(deg)  \end{tabular}} & \multicolumn{3}{c}{HEALPix level 6} & \multicolumn{3}{c}{HEALPix level 12}\\
    \cline{4-6} \cline{7-9}
     & & & $P$ & $\left \langle t_E \right \rangle_\textup{o}$ & $\Gamma$ & $P$ & $\left \langle t_E \right \rangle_\textup{o}$ & $\Gamma$\\
     & & & ($\times 10^{-11}$) & ($\times 10^{-7}$ yr) & ($\times 10^{-5}$ observers/yr/deg$^{2}$) & ($\times 10^{-11}$) & ($\times 10^{-7}$ yr) & ($\times 10^{-5}$ observers/yr/deg$^{2}$)\\
     \hline
    0 & 45.00000 & 0.59684 & 0.490 & 1.71 & 1.83 & 0.356 & 3.99 & 0.569 \\
    1 & 45.70313 & 1.19375 & 0.248 & 1.15 & 1.38 & 0.316 & 3.46 & 0.582 \\
    2 & 44.29688 & 1.19375 & 0.228 & 1.35 & 1.08 & 0.239 & 1.98 & 0.769 \\
    3 & 45.00000 & 1.79078 & 0.317 & 1.75 & 1.15 & 0.255 & 3.86 & 0.421 \\
    4 & 46.40625 & 1.79078 & 0.418 & 1.04 & 2.57 & 0.475 & 3.13 & 0.965 \\
    $\cdot \cdot \cdot$ & $\cdot \cdot \cdot$ & $\cdot \cdot \cdot$ & $\cdot \cdot \cdot$ & $\cdot \cdot \cdot$ & $\cdot \cdot \cdot$ & $\cdot \cdot \cdot$ & $\cdot \cdot \cdot$ & $\cdot \cdot \cdot$ \\
    $\cdot \cdot \cdot$ & $\cdot \cdot \cdot$ & $\cdot \cdot \cdot$ & $\cdot \cdot \cdot$ & $\cdot \cdot \cdot$ & $\cdot \cdot \cdot$ & $\cdot \cdot \cdot$ & $\cdot \cdot \cdot$ & $\cdot \cdot \cdot$ \\
    $\cdot \cdot \cdot$ & $\cdot \cdot \cdot$ & $\cdot \cdot \cdot$ & $\cdot \cdot \cdot$ & $\cdot \cdot \cdot$ & $\cdot \cdot \cdot$ & $\cdot \cdot \cdot$ & $\cdot \cdot \cdot$ & $\cdot \cdot \cdot$ \\
    24573 & 90.70313 & 40.22818   & 24.6 &  2.52 & 62.3 & 25.6 & 3.84   & 42.4 \\
    24574 & 89.29688 & 40.22818   & 33.3 &  3.08 & 68.9 & 33.1 & 6.85   & 30.8 \\
    24575 & 90.00000 & 41.01450   & 22.0 &  2.12 & 66.2 & 21.9 & 4.24   & 32.8 \\
    24576 & 180.00000 & -41.01450 & 1.99 & 0.325 & 39.0 & 2.03 & 0.535  & 24.2 \\
    24577 & 180.70313 & -40.22818 & 1.71 &  1.22 & 8.92 & 1.71 & 2.24   & 4.86 \\
    $\cdot \cdot \cdot$ & $\cdot \cdot \cdot$ & $\cdot \cdot \cdot$ & $\cdot \cdot \cdot$ & $\cdot \cdot \cdot$ & $\cdot \cdot \cdot$ & $\cdot \cdot \cdot$ & $\cdot \cdot \cdot$ & $\cdot \cdot \cdot$ \\
    $\cdot \cdot \cdot$ & $\cdot \cdot \cdot$ & $\cdot \cdot \cdot$ & $\cdot \cdot \cdot$ & $\cdot \cdot \cdot$ & $\cdot \cdot \cdot$ & $\cdot \cdot \cdot$ & $\cdot \cdot \cdot$ & $\cdot \cdot \cdot$ \\
    $\cdot \cdot \cdot$ & $\cdot \cdot \cdot$ & $\cdot \cdot \cdot$ & $\cdot \cdot \cdot$ & $\cdot \cdot \cdot$ & $\cdot \cdot \cdot$ & $\cdot \cdot \cdot$ & $\cdot \cdot \cdot$ & $\cdot \cdot \cdot$ \\
    49147 & 313.59375 & -1.79078 & 1.14  & 0.428 & 16.9 & 1.29  & 0.0948 & 86.9 \\
    49148 & 315.00000 & -1.79078 & 0.536 & 0.613 & 5.57 & 0.392 & 0.616  & 4.06 \\
    49149 & 315.70313 & -1.19375 & 0.538 & 0.643 & 5.33 & 0.602 & 1.57   & 2.44 \\
    49150 & 314.29688 & -1.19375 & 0.923 & 0.673 & 8.73 & 1.03  & 0.799  & 8.17 \\
    49151 & 315.00000 & -0.59684 & 0.719 & 1.31  & 3.51 & 0.614 & 1.91   & 2.05 \\
    \hline
    \end{tabular}
    \begin{flushleft}
    \textbf{Note.} The full table is available in an electronic format.
    \end{flushleft}
\end{table*}

% Don't change these lines
\bsp	% typesetting comment
\label{lastpage}
\end{document}